\documentclass{article}
\usepackage{graphicx}			% To enable including graphics in pdf's
\usepackage{booktabs}			% Better tables
\usepackage{tabulary}			% Support longer table cells
\usepackage[T1]{fontenc}		% Use T1 font encoding for accented characters
\usepackage[utf8]{inputenc}		% For UTF-8 support
\usepackage{xpatch}
\usepackage{xcolor}				% Allow for color (annotations)
\usepackage{listings}			% Allow for source code highlighting
\usepackage[page,toc,title]{appendix}
\usepackage{minitoc}
\usepackage{enumitem}
\usepackage{hyperref}
\usepackage{pdflscape} 
\hypersetup{
   colorlinks=true,
   linkcolor=black,
   citecolor=black,
   filecolor=black,
   urlcolor=black
}
\usepackage{orcidlink}

\usepackage{apacite}
\usepackage[most]{tcolorbox}
\usepackage[most]{tcolorbox}
\usepackage{multibib}
\newcites{SI}{Supplementary Information References}
\newcommand{\SIref}[1]{\hyperref[#1]{Supplementary Information~\ref*{#1}}}
\usepackage{amsthm}

\usepackage{authblk}
\usepackage{float}
\usepackage{tabularx,ragged2e}
\usepackage{longtable}
\usepackage{array}
\usepackage{amsmath}  
\usepackage{pifont}
\newcolumntype{L}[1]{>{\raggedright\let\newline\\\arraybackslash\hspace{0pt}}m{#1}}

\definecolor{lightgray}{gray}{0.97}

\lstdefinestyle{promptstyle}{
  basicstyle=\ttfamily\small,
  backgroundcolor=\color{lightgray},
  breaklines=true,
  columns=fullflexible,
  keepspaces=true,
  showstringspaces=false
}

\newtcblisting{promptbox}{
  listing only,
  colback=lightgray,
  colframe=black!20,
  boxrule=0.3pt,
  arc=2pt,
  left=6pt,
  right=6pt,
  top=6pt,
  bottom=6pt,
  listing options={style=promptstyle}
}

\title{{The Hidden Costs of AI-Mediated Political Outreach: Persuasion and AI Penalties in the US and UK}}
\author[1]{Andreas Jungherr \orcidlink{0000-0003-2598-2453}}
\author[2]{Adrian Rauchfleisch \orcidlink{0000-0003-1232-083X}}
\affil[1]{University of Bamberg}
\affil[2]{National Taiwan University}
\date{\today}

\begin{document}

\maketitle
\begin{abstract}

As AI-enabled systems become available for political campaign outreach, an important question has received little empirical attention: how do people evaluate the communicative practices these systems represent, and what consequences do those evaluations carry? Most research on AI-enabled persuasion examines attitude change under enforced exposure, leaving aside whether people regard AI-mediated outreach as legitimate or not. We address this gap with a preregistered 2×2 experiment conducted in the United States and United Kingdom (N = 1,800 per country) varying outreach intent (informational vs.~persuasive) and type of interaction partner (human vs.~AI-mediated) in the context of political issues that respondents consider highly important. We find consistent evidence for two evaluation penalties. A persuasion penalty emerges across nearly all outcomes in both countries: explicitly persuasive outreach is evaluated as less acceptable, more threatening to personal autonomy, less beneficial, and more damaging to organizational trust than informational outreach, consistent with reactance to perceived threats to attitudinal freedom. An AI penalty is consistent with a distinct mechanism: AI-mediated outreach triggers normative concerns about appropriate communicative agents, producing similarly negative evaluations across five outcomes in both countries. As automated outreach becomes more widespread, how people judge it may matter for democratic communication just as much as whether it changes minds.
\end{abstract}


\section{Introduction}\label{introduction}

The growing technological capabilities, commercial availability, and
cultural normalization of AI-enabled systems have
raised attention toward the uses for AI-enabled persuasion, the
automation of changing people's minds. A growing set of studies shows
that AI can successfully change people's minds be it through the
generation of content for communicative interventions, such as ads or
information treatments
\cite{Bai:2025aa,Chu:2024aa,Goldstein:2024aa,Hackenburg:2024aa,Hackenburg:2023aa,Hackenburg:2025aa,Karinshak:2023aa,Matz:2024aa,Simchon:2024aa,Teeny:2024aa},
or automated dialogue
between people and machines
\cite{Argyle:2025aa,Boissin:2025aa,Chen:2025aa,Chen:2026aa,Costello:2024aa,Costello:2025aa,Costello:2025ab,Costello:2026aa,Crabtree:2025aa,Czarnek:2025aa,Hackenburg:2025ab,Holbling:2025aa,Hopkins:2026aa,Kowal:2025aa,Lin:2025aa,Rand:2025aa,Salvi:2025aa,Schoenegger:2025aa,White:2025aa,Xu:2025aa}.
Given these findings and growing AI use in campaigns in general
\cite{Foos:2024aa,Jungherr:2024aa,neyazi_campaign_2025}, the question is no longer
\emph{if} AI will be used in persuasion attempts but \emph{how} and to
\emph{what effect}. This makes AI-enabled systems an important new
element in persuasion that both academics and practitioners need to
understand better.

Much of the existing empirical literature focuses on persuasion outcomes
for people who are interacting with AI-enabled persuasion in controlled
conditions of survey experiments
\cite{Argyle:2025aa,Boissin:2025aa,Chen:2025aa,Chen:2026aa,Costello:2024aa,Costello:2025aa,Costello:2025ab,Costello:2026aa,Crabtree:2025aa,Czarnek:2025aa,Hackenburg:2025ab,Holbling:2025aa,Hopkins:2026aa,Kowal:2025aa,Lin:2025aa,Rand:2025aa,Salvi:2025aa,Schoenegger:2025aa,White:2025aa,Xu:2025aa}.
Researchers examine whether AI-generated messages change attitudes,
whether conversational AI can persuade people during dialogue, and which
design features or interaction patterns make such interventions more
effective. This research has produced important insights into the
persuasive potential of AI-enabled communication. At the same time, this
research has remained vulnerable to a common challenge in campaign
effects research. Identifying effects of
information treatments under controlled exposure conditions provides
internally valid estimates that do not necessarily generalize to
conditions of real-world exposure (external validity) \cite{Druckman:2012ab}. This
clearly also holds for findings on AI-enabled persuasion
\cite{Chen:2025aa}.

To better understand real-world conditions for AI-enabled persuasion, we
must not only look at persuasion but also address broader questions: How
do people evaluate campaigns that deploy AI-mediated persuasion systems?
And, what consequences do those evaluations have for the legitimacy of
the communicative practice itself and the actors using it? The
large-scale use of automated persuasion systems is constrained by
whether people accept this practice, if they are willing to engage when
they do encounter them, and if exposure does not systematically damage
trust in the organizations responsible. If people evaluate AI-mediated
negatively and as 
damaging to public discourse, the real-world consequences of automated
persuasion may extend well beyond its measurable effects on attitudes.
We therefore shift attention from persuasion outcomes to how people
evaluate anticipated encounters with AI-mediated campaign outreach and
what those evaluations imply for the legitimacy and perceived
acceptability of these communicative practices.

This question is particularly relevant in political communication.
Campaign outreach tries to engage people in interactions
(e.g.~conversations with canvassers, online exchanges, or chatbot
dialogues) as a necessary precondition for exposure to persuasive
messages. Whether people engage, and their evaluation of the encounter
depend substantially on expectations about the interaction itself: what
it is trying to achieve, and who is conducting it. Two features of
campaign outreach are especially consequential for these expectations. First, campaign outreach can take different forms: some focus on information presentation and exchange, while others directly aim at persuasion. Second, advances in generative AI have made it possible to conduct such outreach through automated conversational systems rather than human campaigners.

We argue that these features shape evaluations of anticipated campaign
encounters through related but distinguishable processes. One plausible
interpretation is that explicit persuasive intent is likely to
activate~reactance, as people may perceive a threat to their freedom to
hold their current opinions and therefore evaluate the encounter more
negatively and become less willing to engage. AI-mediated outreach, by
contrast, is likely to trigger~normative judgments about the
appropriateness of agents involved in political communication.

These processes may ultimately shape a broader evaluative stance toward
the anticipated encounter. People form judgments about whether such
interactions are acceptable and worth engaging in. Negative
evaluations may therefore show not only in reduced willingness to
participate but also in perceptions that the outreach threatens personal
autonomy, harms public discourse, or reflects poorly on the
organizations responsible for it.

To examine these dynamics, we conduct a preregistered 2×2 experiment in
the US (N = 1,800) and the United Kingdom (N = 1,800) that varies two
features of campaign outreach on political issues that people consider
highly important: whether the interaction is framed as informational or
persuasive and whether the exchange partner is described as a human
campaigner or an AI chatbot. We measure willingness to participate,
perceived threat to autonomy, acceptability of the outreach, perceived
positive impact, intentions to avoid future campaign contact, and
evaluations of sponsoring organizations.

Running the experiment in the US and the UK provides evidence from two
countries where most prior AI-persuasion experiments have been conducted
and that differ in political systems and campaigning cultures. This
variance allows us to speak about the impact of AI-enabled persuasion
attempts on campaigning more broadly than, for example, focusing
exclusively on the US.

In measuring evaluative responses to anticipated rather than directly
experienced AI-mediated outreach, our design reflects two realistic
conditions under which people encounter AI-enabled political persuasion.
First, the majority of people will encounter AI-enabled persuasion
systems not by interacting with them directly but through media
coverage and public discourse. The society-wide evaluative and normative consequences of
automated persuasion are likely to work primarily through 
mediated awareness rather than through direct interaction. Our
experiment is designed to approximate this condition: how people
evaluate and respond to the prospect of AI-mediated outreach before and
independently of any direct encounter with it. Second, the explicit
disclosure of both AI mediation and persuasive intent in our vignettes
reflects an emerging regulatory stance. Regulatory frameworks in the US,
EU, and UK are moving toward mandatory disclosure of AI-generated
political content and AI use
\cite{Eisert:2025aa,European-Parliament:2024aa,EU:2024ab}.
Accordingly, people will increasingly encounter digital campaign
outreach under conditions of explicit disclosure. Testing both the role
of persuasive intent and AI mediation under full disclosure, therefore,
captures the conditions that policymakers are actively constructing and
that people are increasingly likely to face.

Our results reveal two systematic patterns. First, we observe a strong
persuasion penalty: outreach framed as explicitly persuasive is
evaluated more negatively across five of six outcomes in the US and all
six in the UK, including perceptions of autonomy threat, acceptability,
positive impact, future campaign avoidance, and organizational trust.
Second, we find consistent evidence for an AI penalty: AI-mediated
outreach produces less favorable evaluations on the same five evaluative
outcomes in both countries. These findings suggest that the societal
consequences of AI-enabled campaign outreach cannot be assessed by
examining persuasion effects alone: our findings indicate that when
people become aware of AI-mediated campaign outreach, they form negative
evaluative judgments about its legitimacy that carry downstream
consequences for trust in campaigns and institutions.

More broadly, our findings highlight an important constraint on the
real-world scalability of automated persuasion. The effectiveness of
AI-enabled persuasion cannot be assessed solely by examining whether
AI-generated messages change attitudes once people are exposed to them.
It also depends on whether people are willing to enter such
interactions and whether the practices used to initiate them are
perceived as legitimate. If people view AI-mediated persuasion attempts
negatively, automated persuasion may carry broader consequences for
trust in political communication than opinion change, and may face
significant exposure limits.

\section{Theoretical framework}\label{theory}
 
Two features of campaign outreach are likely to shape people's evaluations of anticipated campaign encounters: the intent of the
outreach (informational vs.~explicitly persuasive) and the nature of the
exchange partner (human vs.~AI-mediated). We expect that these features
activate related but distinguishable processes. Explicit persuasive
intent is likely to trigger reactance: people perceive a threat to their
freedom to hold their current opinions and respond by evaluating the
encounter more negatively. AI mediation might trigger normative
judgments about the appropriateness of communicative agents: people may
perceive the use of machines for political persuasion as violating
expectations about how political communication should happen. Together
these processes can be expected to produce negative evaluations of the
anticipated encounter: a persuasion and an AI penalty.

\subsection{Persuasion penalty}\label{persuasion-penalty}

One explicit goal of political campaigns is to change people's minds
\cite{Foos:2018aa}. Research on campaign persuasion effects
documents the conditions under which campaign contact and messaging
shift political opinions
\cite{Broockman:2023aa,Coppock:2022aa,kalla2018}. How people
evaluate the outreach itself (e.g., whether they regard persuasive
tactics as legitimate, acceptable, or appropriate) is a separate
question that largely remains unaddressed. This matters because
evaluative reactions to the practice of persuasion shape whether people
are willing to engage with it, how they assess the organizations
responsible for it, and what broader consequences exposure carries for
trust in political communication. We ask whether signaling persuasive
intent triggers negative evaluations of the anticipated encounter before
any message is received.

Reactance theory suggests how people are expected to respond to explicit
persuasion attempts \cite{Brehm:1981aa}. When people perceive an
attempt to influence or constrain their opinions, they experience a
motivational response aimed at restoring their freedom of choice
\cite{Steindl:2015aa}. Alerting people to a message's persuasive
intent reduces both persuasion and willingness to engage
\cite{Benoit:1998aa,Petty:1979aa}. When people recognize the
strategic nature of a communication -- what Friestad and Wright
\citeA{Friestad:1994aa} term the activation of persuasion knowledge
-- they adopt coping responses including skepticism and disengagement.
These effects grow with the perceived importance of the threatened
freedom \cite{Brehm:1981aa,Dillard:2005aa}. We therefore expect:

\begin{description}[leftmargin=2em, labelwidth=1.5em, font=\normalfont\itshape]
  \item[H1a] (Willingness to participate): People will be less interested in engaging when outreach is persuasive rather than informational.
  \item[H1b] (Perceived threat to freedom): People will rate outreach as a higher threat to freedom when it is persuasive rather than informational.
  \item[H1c] (Acceptability): People will rate outreach as less acceptable when it is persuasive rather than informational.
  \item[H1d] (Positive impact): People will perceive outreach as having less positive impact when it is persuasive rather than informational.
\end{description}

In our study, we focus on issues that are important to people, since much of the AI-enabled persuasion research focuses on
high-salience topics such as immigration
\cite{Argyle:2025aa,Rand:2025aa}, climate change
\cite{Czarnek:2025aa,Remshard:2026aa}, conspiracy beliefs
\cite{Boissin:2025aa,Costello:2024aa,Costello:2025ab,Costello:2026aa,Hopkins:2026aa},
or the correction of misinformation
\cite{DiGiuseppe:2026aa,Goel:2025aa}. In such scenarios, where issues are
regarded as personally important, reactance should be especially strong
\cite{Dillard:2005aa}.

\subsection{AI penalty}\label{ai-penalty}
Independent of persuasive intent, the AI-mediated nature of outreach may
itself produce negative evaluations through a related but distinguishable mechanism. AI-mediated communication, which \citeA{Hancock:2020aa} define as interpersonal communication in which a computational agent modifies, augments, or generates messages on behalf of a communicator, transforms the character of the interaction even when content remains unchanged. Where reactance responds to the
intent of a message, the AI penalty responds to the perceived
appropriateness of the agent delivering it. The structural properties of AI that make it normatively inappropriate may also register as a threat to the conditions under which people form opinions freely, producing some elevation of freedom-threat perceptions through this route even in the absence of explicitly persuasive intent.

Political persuasion carries implicit expectations about authenticity and
accountability. When a human campaigner
contacts a voter, the interaction can be read as expressing genuine
conviction or civic commitment, even when strategically motivated. Field
experiments consistently show that in-person contact is more effective
for mobilization than other interventions \cite{Green:2019ab}, and
deep canvassing (an approach that relies on non-judgmental narrative
exchange and perspective-taking) has been shown to durably shift
entrenched attitudes \cite{Broockman:2016aa,Kalla:2020aa}. One
reason driving these findings is that a human exchange partner can, at
least in principle, be held socially accountable for what they say and
how they say it.

Delegating outreach to a machine disrupts these expectations. Research
consistently documents resistance to algorithmic judgment in domains
perceived as subjective, identity-relevant, or morally charged
\cite{Castelo:2019aa,Dietvorst:2015aa,Longoni:2019aa,Morewedge:2022aa}.
This resistance is strongest when tasks require the sensitivity to
individual circumstances and moral reasoning associated with human minds
\cite{Bigman:2018aa,Cadario:2021aa}. Political opinion formation
clearly shares these characteristics. When people learn that a
communicative agent is a machine, they attribute stereotypically
mechanical properties to it (e.g., impersonality, strategic
optimization, emotional coldness) even when its output is
indistinguishable from human communication
\cite{Sundar:2008aa,Sundar:2020aa,Yan:2024aa}. These attributions
are acceptable in technical contexts where objectivity is valued
\cite{Lee:2018ab}, but conflict with what political communication is
expected to be. Prior work in political contexts directly supports this.
AI involvement in campaign outreach is largely perceived as a norm
violation \cite{Jungherr:2024aa}, and AI facilitation of democratic
deliberation reduces both willingness to participate and evaluations of
expected quality relative to identical human-facilitated formats
\cite{Jungherr:2025aa}. We therefore expect:

\begin{description}[leftmargin=2em, labelwidth=1.5em, font=\normalfont\itshape]
\item[H2a] (Willingness to participate): People will be less interested in
  engaging when outreach is AI-mediated rather than human-mediated.
\item[H2b] (Perceived threat to freedom): People will rate outreach as a
  higher threat to freedom when it is AI-mediated rather than
  human-mediated.
\item[H2c] (Acceptability): People will rate outreach as less acceptable when
  it is AI-mediated rather than human-mediated.
\item[H2d] (Positive impact): People will perceive AI-mediated outreach as
  having less positive impact than human-mediated outreach.
\end{description}

\subsection{Downstream effects}\label{downstream-effects}

Both reactance and perceived norm violation can extend beyond the
immediate evaluative response, shaping how individuals view the
organizations responsible for the outreach. Research on persuasion
knowledge shows that recognizing influence attempts prompts negative
updating of the persuading agent
\cite{Friestad:1994aa,Campbell:2000ab}, and research on campaign
communication shows that objectionable practices can depress broader
political engagement \cite{Fridkin:2011aa,Lau:2007aa}. We therefore
predict that both persuasive framing and AI mediation will produce
downstream consequences beyond the immediate encounter:

\begin{description}[leftmargin=2em, labelwidth=1.5em, font=\normalfont\itshape]
\item[H1e] (Future campaign avoidance): People will report greater future
  avoidance when campaign outreach is persuasive rather than
  informational.
\item[H1f] (Penalty for source): People will report more negative evaluations
  of associated organizations when campaign outreach is persuasive
  rather than informational.
\item[H2e] (Future campaign avoidance): People will report greater future
  avoidance when campaign outreach is AI-mediated rather than
  human-mediated.
\item[H2f] (Penalty for source): People will report more negative evaluations
  of associated organizations when campaign outreach is AI-mediated
  rather than human-mediated.
\end{description}

\subsection{Interaction: how persuasive intent and AI mediation
combine}\label{interaction-how-persuasive-intent-and-ai-mediation-combine}

While we assume that the persuasion penalty and the AI penalty follow
from distinguishable processes, they need not operate independently. The
properties that make AI mediation normatively objectionable (e.g.,
perceived strategic optimization, scalability, and absence of social
accountability) also bear directly on how threatening a persuasive
attempt feels. A human campaigner who tries to persuade operates within
a framework of mutual social obligation; a machine that tries to
persuade may be perceived as an instrument of asymmetric, frictionless
manipulation with no social skin in the game. AI mediation may therefore
not merely add a normative penalty on top of reactance but actively
intensify it, making the encounter feel simultaneously more threatening
and more normatively inappropriate. Under this amplification logic, the
combination of persuasive intent and AI mediation should produce
reactions stronger than either feature would generate alone.

A competing prediction follows from a different theoretical starting
point. Research on impression formation and evaluative judgment
consistently finds that people integrate multiple pieces of information
through averaging rather than addition, meaning that each successive
piece of same-valenced information produces a smaller marginal shift in
the overall evaluation than the one before it
\cite{Anderson:1971aa}. This averaging dynamic is reinforced
diminishing sensitivity: the subjective impact of negative features
decreases as evaluations move further from a neutral reference point,
comparable to patterns known from decision-making and judgment research
\cite{Kahneman:1979aa,Tversky:1992aa}. Under this attenuation logic,
if persuasive intent already triggers strong reactance and drives
evaluations substantially downward, AI mediation has limited room to
worsen them further and vice versa. The two penalties would then
substitute for each other rather than compound, with each successive
negative feature contributing less to the overall evaluation than it
would have in isolation.

Both predictions are theoretically coherent but draw on different
arguments: amplification rests on the specific ways AI mediation makes
persuasive intent appear more calculated; attenuation rests on
structural properties of evaluative judgment that apply regardless of
mechanism content. We preregistered the amplification prediction on the
grounds that the conceptual link between the two mechanisms was
specifically reinforcing rather than merely additive. The properties of
AI that may trigger normative concern are precisely the properties that
make a persuasive attempt feel harder to resist. Attenuation, though
plausible as a general feature of evaluative judgment, required no
specific assumptions about these mechanisms in particular.

\begin{description}[leftmargin=2em, labelwidth=1.5em, font=\normalfont\itshape]
\item[H3] (Interaction effect): For each outcome (a--f), the negative effect
  of persuasive (vs.~informational) outreach will be stronger when
  outreach is AI-mediated (vs.~human-mediated).
\end{description}

Running the experiment in the US and UK provides an opportunity to
examine whether this interaction varies with the political communication
environment. US campaigns operate across longer electoral cycles with
higher contact frequency and less regulated spending
\cite{Sides:2023aa}; UK campaigns are more compressed and subject to
stricter broadcast and expenditure rules \cite{Ford:2025aa}. These
differences may shape the baseline intensity of reactance that
persuasive intent produces. If persuasive outreach is less normalized in
the UK and a less routine feature of political life, it may be evaluated
more categorically, leaving less evaluative space for AI mediation to
add further deterioration and producing attenuation where the US shows
more independent additive effects.

\subsection{Moderators}\label{moderators}

Individual differences in how people evaluate AI-mediated outreach are
likely to shape the magnitude of both the persuasion penalty and the AI
penalty. We examine three individual-level moderators that bear on
people's openness to political contact and their prior orientations
toward AI.

First, people who habitually avoid political conversations may respond
more negatively to any form of campaign outreach, but may be especially
sensitive to AI-mediated contact, which removes the social cues and
interpersonal accountability that can make human conversations feel
manageable \cite{Boland:2024aa}. Second, feelings toward people with
opposing opinions capture the degree of affective polarization that
people bring to anticipated encounters with campaign outreach
\cite{Lelkes:2017aa}. Those with more negative feelings toward
political opponents may evaluate any outreach from an opposing campaign
more harshly, thereby amplifying both the persuasion and AI penalties.
Third, general AI risk perceptions should be especially relevant to
evaluations of AI-mediated outreach specifically: people who perceive AI
as threatening or risky are more likely to read AI involvement in
political communication as a normative violation, amplifying the AI
penalty \cite{Jungherr:2025aa}. While prior research suggests that
these factors matter in related contexts
\cite{Jungherr:2025aa,Jungherr:2024aa}, theory provides limited
guidance on the strength of these effects in our setting. We therefore
formulate these expectations as research questions rather than
hypotheses:

\begin{description}[leftmargin=2em, labelwidth=1.5em, font=\normalfont\itshape]
\item[RQ1] Does avoidance of political conversations moderate the effect of
  AI-mediated outreach?
\item[RQ2] Does the feeling toward people with opposing opinions moderate
  the effect of AI-mediated outreach?
\item[RQ3] Does the general AI risk perception moderate the effect of
  AI-mediated outreach?
\end{description}

\section{Materials \& Methods}\label{materials-methods}

To test the preregistered hypotheses,\footnote{Preregistration US:
  \url{https://osf.io/sjh3v/overview?view_only=a6e7ef77368843bd9cdca37904712114}
  ; Preregistration UK:
  \url{https://osf.io/nceht/overview?view_only=93038c0124ee431fb3de6dec8bb95acc}. The data and code to reproduce the findings reported in this study are available on OSF: \url{https://osf.io/zvr7f/overview?view_only=21c8e4c35d3642f79dfa64d1b9d55d87}}
we conducted two parallel 2 (outreach mode: human vs.~AI-mediated) × 2
(outreach intent: informational vs.~persuasive) between-subjects
experiments, one in the United States and one in the United Kingdom,
both using the Prolific platform. In each country, we set a quota of
1,800 completion ($\approx$450 per condition), based on a power analysis
conducted prior to the experiment and documented in the preregistration,
and stopped data collection once the quota was reached. The design was reviewed and approved  by the IRB of the University of Bamberg.

The US study ran from 6 March to 14 March 2026; the UK study ran from 9
February to 20 February 2026. Following preregistered criteria, we
excluded participants who failed two attention checks prior to treatment
exposure; these were directly excluded from data collection and did not
count toward the total of 1,800 completes per country. We applied
quota-based sampling to approximate each country's population by age,
gender, and party identification (see \SIref{section:pop} for sociodemographic distributions and data-quality procedures).

The study design, questionnaire structure, treatments, and measures were
identical across both countries. After participants provided informed
consent, we began each survey with questions about sociodemographics and
political orientation, followed by our preregistered moderators:
avoidance of political conversations (four items, adapted from \citeNP{Boland:2024aa}) and general AI risk perceptions (four items; \citeNP{Jungherr:2025aa}), both showing good internal consistency in
both samples (see Table 1). To prevent negative priming of AI, we also
included items asking about AI benefits, which were not used for further
analysis. Before participants viewed one of the four vignettes, we asked
them to think about ``one political or social issue that matters to you
personally'' and note it in a free-text field. To assess ecological
validity, we also asked them to rate how important the issue is to them
(1 = Not at all important; 7 = Extremely important). This research
design choice was intentional: based on pretesting, we expected
participants to reliably select issues they regarded as personally
important, allowing us to reduce variation in baseline issue importance
across respondents while increasing experimental realism by focusing on
issues that matter to them. Indeed, issue importance ratings were high
in both countries (US: M = 6.46, SD = 0.79; UK: M = 6.22, SD = 0.90),
indicating that participants generally selected issues they cared about
deeply. The issue profiles also broadly matched each country's political
context and partisan cleavages (see \SIref{section:issues}).~
Furthermore, in both countries, fewer than 1\% of participants selected
a value below the scale midpoint. We used this approach to study
campaign outreach in a context close to that of much of the existing
AI-enabled persuasion literature, which has largely focused on
high-salience issues. We also asked respondents about their feelings
toward people with opposing opinions \cite{Lelkes:2017aa}.

Participants were then randomly assigned to one of the four treatments and
required to remain on the page for at least 10 seconds (seconds on page
US: M = 21.66, SD = 25.98; UK: M = 20.25, SD = 26.30). In the different
conditions, only the outreach mode and outreach intent were varied; all
other wording was held constant, stating that the campaign behind the
outreach ``takes a position that is different from yours'' (see
\SIref{section:treatments} for the full treatment texts). We then
measured willingness to participate in such outreach, followed by
acceptability, perceived positive impact, future campaign avoidance,
source penalty, and perceived threat to freedom
\cite{Dillard:2005aa}. Except for willingness to participate (a
single item) and source penalty (four items), all outcomes were measured
with three items each and showed good internal consistency in both
samples (see Table 1; full item wordings and item-level descriptive statistics are provided in \SIref{section:items}). At the very end, we also included a few questions
about the treatment content, confirming that participants received the
treatments (see \SIref{section:manipchecks} for manipulation checks
and instrumental variable robustness checks). 

We tested hypotheses using OLS regression models with effect coding for
both factors (-0.5, +0.5) and their interaction. With this coding
scheme, the main-effect coefficients represent the mean difference
between the two levels of each factor, averaged over the other factor.
We estimated models separately for each country. For the research
questions, we first centred the moderators and added them as moderators
for the outreach mode variable (human vs.~AI-mediated) in an OLS
regression model.

\begin{table}[H]
\begin{tabular}{@{}p{3.6cm}p{3.2cm}p{1cm}p{1cm}p{1cm}@{}}
\toprule
Variable & $\alpha$ & US M (SD) & UK M (SD) & $n$ \\
\midrule
Outcome variables & & & & \\

H1/2a: Willingness to participate & --- & 3.58 (2.05) & 3.80 (1.90) & 1,800 \\
H1/2b: Perceived threat to freedom & US = .86 / UK = .87 & 3.61 (1.70) & 3.81 (1.59) & 1,800 \\
H1/2c: Acceptability & US = .79 / UK = .80 & 3.63 (1.51) & 3.59 (1.38) & 1,800 \\
H1/2d: Perceived positive impact & US = .84 / UK = .84 & 3.70 (1.59) & 3.87 (1.42) & 1,800 \\
H1/2e: Future campaign avoidance & US = .90 / UK = .90 & 4.51 (1.71) & 4.50 (1.59) & 1,800 \\
H1/2f: Penalty for source & US = .84 / UK = .84 & 4.38 (1.36) & 4.53 (1.24) & 1,800 \\
Moderator variables & & & & \\
RQ1: Avoidance of political conversations & US = .88 / UK = .86 & 4.39 (1.81) & 3.84 (1.66) & 1,800 \\
RQ2: Feelings toward people with opposing opinions & --- & 35.71 (29.19) & 36.55 (25.24) & 1,800 \\
RQ3: AI risk perception & US = .82 / UK = .81 & 5.15 (1.32) & 5.07 (1.16) & 1,800 \\
\bottomrule
\end{tabular}
\caption{Descriptive Statistics for All Outcome and Moderator Variables. Almost all outcome items used 7-point scales (1 = Strongly disagree / Not at all interested; 7 = Strongly agree / Very interested). Feelings toward others used a 0--100 scale. All internal consistency coefficients are Cronbach's $\alpha$. $n = 1{,}800$ per country.}
\end{table}

\section{Results}\label{results}

\begin{figure}[H]
    \centering
    \includegraphics[width=\linewidth]{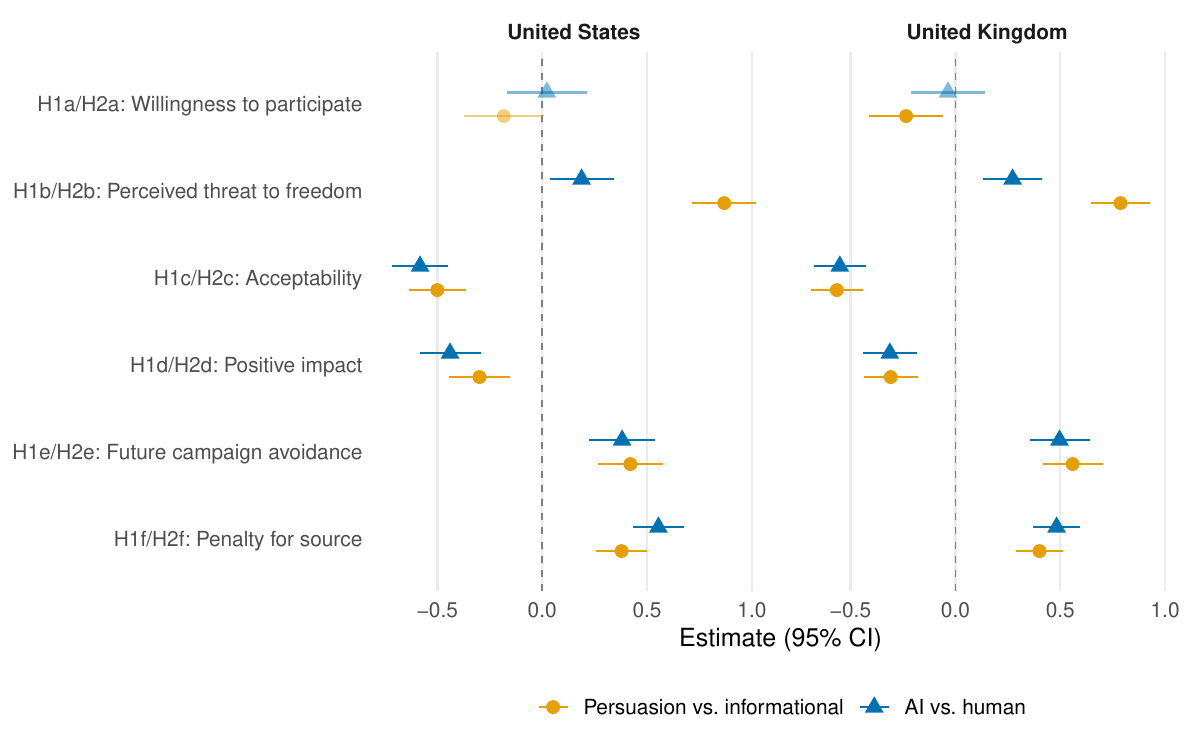}
    \caption{Estimated effects of persuasive outreach (H1, orange) and AI-mediated outreach (H2, blue) on six outcome variables, for the US (left) and the UK (right). Points are OLS regression coefficients with 95\% confidence intervals. The coefficients represent the mean difference between the two levels of each factor, averaged over the other factor. Non-significant estimates are transparent.}
    \label{fig:forest}
\end{figure}

We first analyzed the preregistered persuasion penalty across all
outcome variables (H1a--H1f), comparing persuasive with informational
outreach, followed by the AI penalty across the same outcomes
(H2a--H2f), comparing AI-mediated with human-mediated outreach. We then
analyzed the interaction hypothesis (H3), namely, whether the negative
effect of persuasive (vs.~informational) outreach would be amplified
when outreach was AI-mediated rather than human-mediated. Finally, we
examined three preregistered research questions about individual-level
moderators of the AI penalty: avoidance of political conversations
(RQ1), feelings toward people with opposing opinions (RQ2), and AI risk
perception (RQ3). We report results for the US and UK in parallel
throughout and explicitly note where findings diverge. Complete model
tables are reported in \SIref{section:models}. The preregistered
main effects for the persuasion penalty (H1) and AI penalty (H2) across
outcomes and countries are summarized in Figure~\ref{fig:forest}.

\subsection{Persuasion penalty}\label{persuasion-penalty-1}

In the US, results for the persuasion penalty were largely consistent
with H1b--H1f; only the willingness to participate (H1a) was not
significant. Persuasive outreach did not significantly reduce
willingness to participate in the US (b = -.18, p = .059, 95\% CI
{[}-.37, .01{]}). In the UK, however, all persuasion-penalty hypotheses
were supported, including willingness to participate (b = -.23, p =
.009, 95\% CI {[}-.41, -.06{]}).

For the remaining outcomes, the pattern was the same in both countries.
H1b was supported: persuasive outreach increased perceived threat to
freedom in both the US (b = .87, p \textless{} .001, 95\% CI {[}.72,
1.02{]}) and the UK (b = .79, p \textless{} .001, 95\% CI {[}.64,
.93{]}). In line with H1c, persuasive outreach reduced perceived
acceptability in both countries (US: b = -.50, p \textless{} .001, 95\%
CI {[}-.63, -.37{]}; UK: b = -.57, p \textless{} .001, 95\% CI {[}-.69,
-.44{]}). Our data also supported H1d, as persuasive outreach reduced
perceived positive impact in both the US (b = -.30, p \textless{} .001,
95\% CI {[}-.44, -.15{]}) and the UK (b = -.31, p \textless{} .001, 95\%
CI {[}-.44, -.18{]}). The two hypotheses concerning downstream negative
reactions were likewise supported. In line with H1e, persuasive outreach
increased future campaign avoidance in both the US (b = .42, p
\textless{} .001, 95\% CI {[}.26, .58{]}) and the UK (b = .56, p
\textless{} .001, 95\% CI {[}.42, .70{]}). Consistent with H1f,
persuasive outreach increased negative evaluations of the associated
organization in both the US (b = .38, p \textless{} .001, 95\% CI
{[}.26, .50{]}) and the UK (b = .40, p \textless{} .001, 95\% CI {[}.29,
.51{]}).

Overall, the persuasion-penalty hypotheses found consistent support
across both countries and five of the six outcome variables. The one
exception was the willingness to participate in the US, which was not
significant.

\subsection{AI penalty}\label{ai-penalty-1}

Results for the AI penalty were highly consistent across the two
countries. In neither the US nor the UK did AI-mediated outreach
significantly affect willingness to participate, providing no support
for H2a in either sample (US: b = .02, p = .820, 95\% CI {[}-.17,
.21{]}; UK: b = -.03, p = .696, 95\% CI {[}-.21, .14{]}).

\hspace{0pt}Supporting H2b, AI-mediated outreach increased perceived
threat to freedom in both the US (b = .19, p = .016, 95\% CI {[}.04,
.34{]}) and the UK (b = .27, p \textless{} .001, 95\% CI {[}.13,
.42{]}). Supporting H2c, AI-mediated outreach reduced acceptability in
both the US (b = -.58, p \textless{} .001, 95\% CI {[}-.72, -.45{]}) and
the UK (b = -.55, p \textless{} .001, 95\% CI {[}-.67, -.43{]}).
Supporting H2d, AI-mediated outreach also reduced perceived positive
impact in both the US (b = -.44, p \textless{} .001, 95\% CI {[}-.58,
-.29{]}) and the UK (b = -.31, p \textless{} .001, 95\% CI {[}-.44,
-.18{]}). In line with H2e, AI-mediated outreach increased future
campaign avoidance in both the US (b = .38, p \textless{} .001, 95\% CI
{[}.23, .54{]}) and the UK (b = .50, p \textless{} .001, 95\% CI {[}.35,
.64{]}). Consistent with H2f, AI-mediated outreach increased negative
evaluations of the associated organization in both the US (b = .55, p
\textless{} .001, 95\% CI {[}.43, .68{]}) and the UK (b = .48, p
\textless{} .001, 95\% CI {[}.37, .59{]}).

Thus, the AI penalty was supported for five of the six outcomes in both
countries, with willingness to participate as the consistent exception.

\subsection{Interactions}\label{interactions}

We next tested whether the negative effect of persuasive
(vs.~informational) outreach was amplified when outreach was AI-mediated
(H3). Contrary to H3, significant interaction effects, when present,
indicated attenuation rather than amplification of the persuasion
penalty under AI mediation. Furthermore, the pattern of results differed
considerably between the two countries, with no significant results in
the US but four interactions indicating attenuation in the UK.

For willingness to participate, the interaction was significant in the
UK (b = .40, p = .025, 95\% CI {[}.05, .75{]}), indicating that,
contrary to H3, the persuasion penalty on willingness was attenuated
rather than amplified under AI-mediated outreach. In the US, the
interaction was not significant (b = .25, p = .202, 95\% CI {[}-.13,
.63{]}). For perceived threat to freedom, neither country yielded a
significant interaction (US: b = -.17, p = .269, 95\% CI {[}-.48,
.13{]}; UK: b = -.25, p = .083, 95\% CI {[}-.54, .03{]}). The
interaction was also not significant for acceptability in either country
(US: b = .20, p = .148, 95\% CI {[}-.07, .47{]}; UK: b = .21, p = .101,
95\% CI {[}-.04, .45{]}).

\begin{figure}[H]
    \centering
    \includegraphics[width=\linewidth]{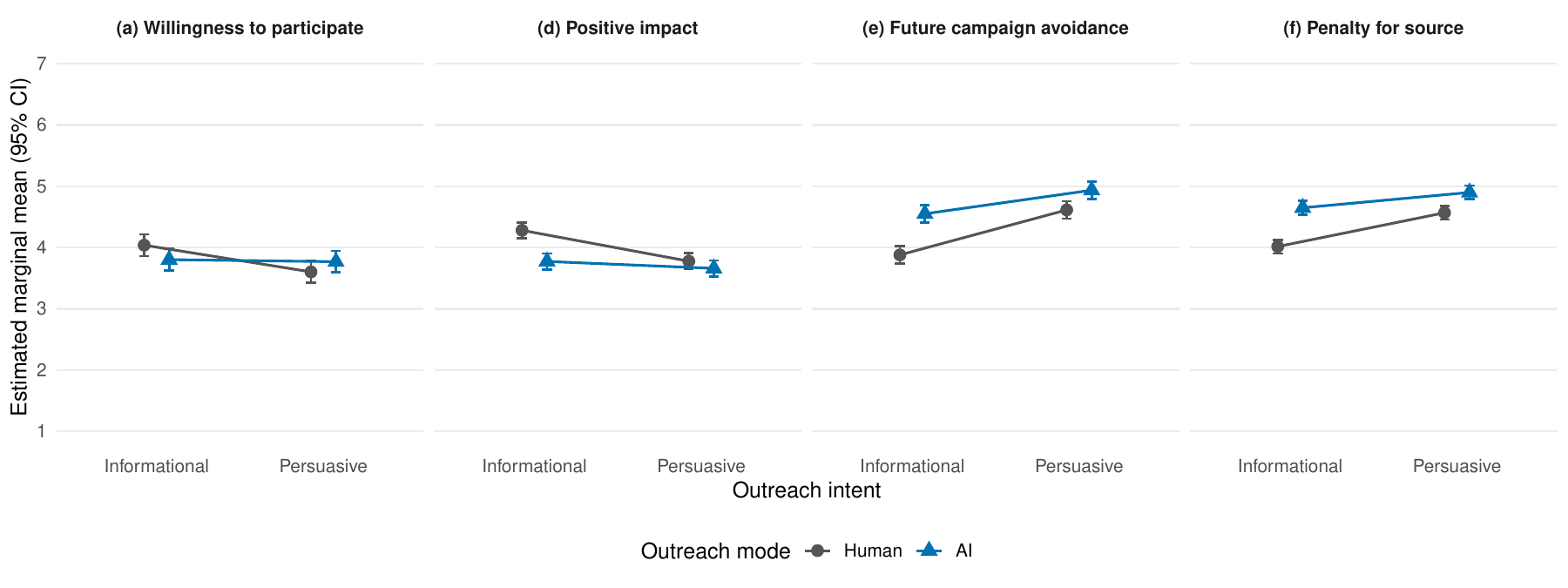}
    \caption{Estimated marginal means for four outcomes with significant interaction effects in the United Kingdom, by outreach intent (informational vs. persuasive) and outreach mode (human vs. AI). Error bars represent 95\% confidence intervals.}
    \label{fig:interactions}
\end{figure}

For perceived positive impact, the two countries diverged. In the UK,
the interaction was significant (b = .39, p = .003, 95\% CI {[}.13,
.65{]}), again reflecting attenuation: the AI penalty on positive impact
was smaller in the persuasive condition than in the informational
condition. In the US, the interaction was not significant (b = .08, p =
.600, 95\% CI {[}-.21, .37{]}). We observed a similar divergence for
future campaign avoidance: the interaction was significant in the UK (b
= -.35, p = .017, 95\% CI {[}-.64, -.06{]}) but was not in the US (b =
-.27, p = .093, 95\% CI {[}-.58, .04{]}). Finally, for the penalty for
source, the interaction was significant in the UK (b = -.30, p = .008,
95\% CI {[}-.52, -.08{]}) but not in the US (b = -.13, p = .293, 95\% CI
{[}-.38, .11{]}). These UK-specific interaction patterns are visualized
in Figure~\ref{fig:interactions} for willingness to participate, positive impact, future
campaign avoidance, and penalty for source.

\hspace{0pt}The amplification hypothesis found no support in either
sample. In fact, the significant UK interactions consistently indicated
the reverse: persuasive outreach attenuated rather than amplified the AI
penalty.

To better understand these patterns, we conducted preregistered planned simple-effects contrasts with Holm-adjusted p-values, comparing AI- versus human-mediated outreach separately within the informational and persuasive conditions (see \SIref{section:contrasts} for the full set of contrasts). Within the informational outreach condition, AI-mediated outreach generally yielded less favorable evaluations than human-mediated outreach across nearly all outcomes in both countries, with the sole exception of willingness to participate. Within the persuasive outreach condition, the AI--human gap became smaller on several measures, especially in the UK, while remaining significant for acceptability, future campaign avoidance, and penalty for source. The clearest cross-national difference concerned perceived positive impact: in the US, AI-mediated persuasive outreach still yielded significantly less favorable evaluations than human-mediated persuasive outreach, whereas in the UK this difference was no longer significant. At the same time, the observed attenuation for some of the variables in the UK did not mean that the AI penalty disappeared: within the persuasive condition, AI-mediated outreach was still evaluated significantly less favorably than human-mediated outreach for two of the four outcomes that showed attenuation, namely future campaign avoidance and source penalty.

\subsection{Moderators}\label{moderators-1}

We also preregistered three moderator variables and tested their
interactions with the AI-versus-human outreach factor. Full model tables
are reported \SIref{section:moderators}

\subsubsection{RQ1: Avoidance of Political
Conversations}\label{rq1-avoidance-of-political-conversations}

Results for avoidance as a moderator diverged substantially between the
two countries. In the UK, avoidance of political conversations
significantly moderated the AI penalty on willingness to participate (b
= .15, p = .004, 95\% CI {[}.05, .26{]}): participants with higher
avoidance showed a more positive response to AI-mediated compared to
human-mediated outreach. In contrast, this interaction was not
significant in the US (b = .10, p = .070, 95\% CI {[}-.01, .20{]}).

For perceived threat to freedom, the moderation pattern pointed in the
opposite direction in the UK: higher avoidance was associated with a
smaller AI penalty on perceived threat (b = -.09, p = .034, 95\% CI
{[}-.18, -.01{]}). In the US, this interaction was not significant (b =
.02, p = .655). For all other outcomes (acceptability, positive impact,
future avoidance, penalty for source), the moderation by avoidance was
not significant in either country.

\subsubsection{RQ2: Feelings Toward People with Opposing
Opinions}\label{rq2-feelings-toward-people-with-opposing-opinions}

The moderating role of outgroup feelings showed the opposite
cross-national pattern from avoidance: significant moderation occurred
in the US but not in the UK for most outcomes. In the US, more positive
feelings toward people with opposing opinions were associated with a
smaller AI penalty on acceptability (b = .01, p = .020), positive impact
(b = .01, p = .033), future campaign avoidance (b = -.01, p = .037), and
penalty for source (b = -.01, p \textless{} .001). In the UK, none
reached significance. For willingness to participate and perceived
threat to freedom, outgroup feelings did not significantly moderate the
AI effect in either country.

\subsubsection{RQ3: AI Risk Perception}\label{rq3-ai-risk-perception}

\begin{figure}[H]
    \centering
    \includegraphics[width=0.75\linewidth]{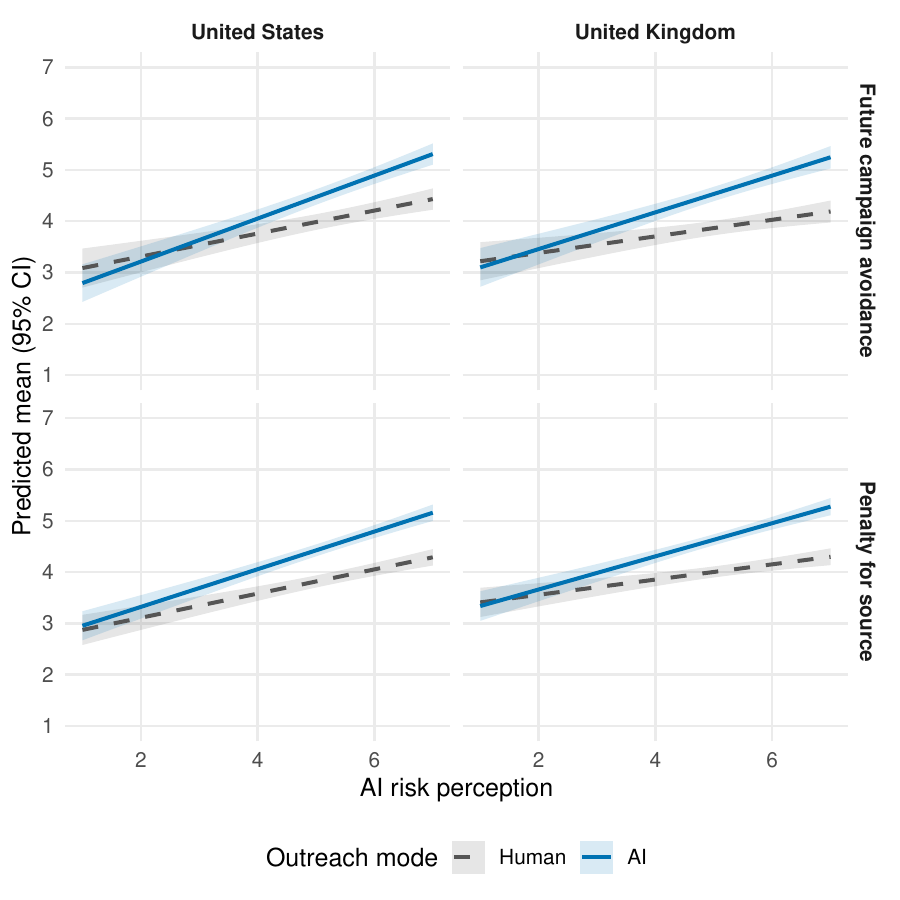}
    \caption{Predicted means for future campaign avoidance (top) and penalty for source (bottom) as a function of AI risk perception, separately for human-mediated (dashed) and AI-mediated (solid) outreach, in the United States (left) and the United Kingdom (right). Shaded bands represent 95\% confidence intervals. All four interactions are statistically significant}
    \label{fig:moderators}
\end{figure}

Higher AI risk perception moderated the AI penalty in both countries,
though the pattern of significant outcomes differed. In the US, higher
AI risk perception amplified the AI effect on perceived threat to
freedom (b = .14, p = .019), acceptability (b = -.14, p = .006), future
campaign avoidance (b = .20, p \textless{} .001), and penalty for source
(b = .13, p = .004). For willingness to participate, the interaction
indicated that higher AI risk perception was associated with a larger AI
penalty (b = -.15, p = .034). In the UK, the interactions for
willingness and perceived threat to freedom did not reach significance
(b = -.09, p = .260 and b = .08, p = .196, respectively), while the
interactions for acceptability (b = -.13, p = .017), positive impact (b
= -.14, p = .015), future campaign avoidance (b = .20, p = .001), and
penalty for source (b = .17, p \textless{} .001) were significant.
Figure~\ref{fig:moderators} visualizes this moderation pattern for future campaign
avoidance and penalty for source, the two outcomes for which AI risk
perception most consistently amplified the AI penalty across both
countries.

Across both countries, AI risk perception consistently amplified the AI
penalty on future campaign avoidance and penalty for source, suggesting
these negative reactions to AI-mediated outreach are especially strong
among individuals who already perceive AI as risky. The US showed an
additional significant moderation on willingness and perceived threat to
freedom that did not replicate in the UK.

\section{Discussion}\label{discussion}

Our findings reveal two systematic patterns in how people evaluate
anticipated encounters with campaign outreach. Both the intent of the
outreach and the use of AI mediation shift evaluations in a negative
direction.

The persuasion penalty is strong, consistent, and the results are
broadly consistent with reactance theory. Outreach framed as explicitly
persuasive is evaluated more negatively across nearly all outcomes in
both countries, including perceptions of autonomy threat, acceptability,
positive impact, future campaign avoidance, and organizational trust.
When individuals anticipate an interaction designed to challenge an
attitude they hold personally important, they respond by evaluating the
encounter as less legitimate and less worth engaging with before any
persuasive message has been received.

The AI penalty is similarly consistent. AI-mediated outreach produces
less favorable evaluations on five outcomes in both countries
(i.e.~perceived autonomy threat, acceptability, perceived positive
impact, future campaign avoidance, and organizational trust) consistent
with the mechanism we proposed: delegating political persuasion to
machines appears to trigger normative concerns about appropriate
communicative agents that manifest as negative assessments of the
practice's legitimacy and the trustworthiness of organizations that use
it. To further explore the norm violation argument, we conducted an additional non-preregistered analysis with the item most directly capturing normative appropriateness ("This kind of outreach is not how campaigns should operate"). The result provides further support for this interpretation: the AI penalty on this item exceeds the persuasion penalty in both countries, while the reverse holds for perceived threat to freedom (see \SIref{section:norm}).

While we could not identify comparably consistent differences in
willingness to participate, except in the UK under persuasive intent,
this pattern likely reflects the context of political issues that
citizens consider highly important (see \SIref{section:issues} for
a content analysis of the issues mentioned). Across both countries, the
mean willingness to participate was below the scale midpoint of 4 in all
conditions except one in the UK (human-mediated informational). Many of
the mentioned issues were highly contentious, suggesting a general
reluctance to engage in any form of interaction with representatives,
human or chatbot, who promote opposing opinions. This interpretation is
supported by an analysis of the issues mentioned, which shows that
respondents mostly cited issues central to contemporary political
conflict in each country (see \SIref{section:issues}). In such a
setting, baseline willingness to engage may already be relatively low,
leaving little room for further declines across conditions. Mean
willingness to participate was also somewhat lower in the US (3.58) than
in the UK (3.80), indicating a modest cross-national difference in
overall readiness to engage. It also has a practical implication that is
easy to miss: campaigns deploying AI-mediated outreach may still secure
participation while simultaneously degrading the evaluative environment
in which that engagement occurs. Participation metrics alone may not
reveal the normative costs reflected in perceptions of legitimacy and
organizational trust.

Contrary to our preregistered prediction, the interaction between
persuasive intent and AI mediation produced attenuation rather than
amplification, and only in the UK. Where the two mechanisms were
expected to reinforce each other, the data show instead that the
persuasion penalty and AI penalty partially substitute for each other:
when one mechanism has already driven evaluations substantially
downward, the second has less room to add further deterioration. The
non-significant results for all interactions in the US context suggest
the two penalties operate more independently in that context,
contributing separate additive shifts without either compounding or
attenuating. This cross-national difference is consistent with a
theoretically meaningful difference in how categorically citizens
evaluate persuasive political outreach: US campaigns operate across
longer electoral cycles with higher contact frequency, making persuasive
outreach a more normalized, if disliked, feature of political life
\cite{Sides:2023aa}. UK campaigns are more compressed and more
tightly regulated, and persuasive outreach may therefore feel more
categorically at odds with what political communication normally looks
like \cite{Ford:2025aa}. One possible interpretation is that under a
more categorical rejection, the evaluative floor is lower, and the AI
penalty has less space to operate. The boundary condition implied by
this reasoning that attenuation emerges when initial reactance is
strong, and amplification may emerge when it is weaker, is a promising
area for future comparative research. A related cross-national contrast also appears in a supplementary non-preregistered analysis: political orientation more clearly moderates reactions to persuasive outreach in the UK, whereas in the US it more clearly moderates reactions to AI-mediated outreach (Appendix~\ref{section:polor}).

A strength of our design is that we study the impact of AI-enabled
persuasion under conditions of explicit disclosure. This reflects an
emerging regulatory direction. Across multiple jurisdictions, including
the US and EU
\cite{Eisert:2025aa,European-Parliament:2024aa,EU:2024ab},
policymakers are constructing frameworks requiring citizens to be
informed when they encounter AI in political contexts. These frameworks
converge on a regulatory environment in which people will increasingly
be informed if interacting with an AI in political contexts. Our
manipulation, in which both exchange partner type and outreach intent
are explicitly stated, corresponds with the disclosure conditions these
frameworks are designed to produce and constitutes a direct test of
citizen responses to AI-mediated political communication under the
transparency norms policymakers are actively establishing.

The practical implications of these findings are sharpest for
informational rather than explicitly persuasive outreach. The AI penalty
on evaluative outcomes is largest when the interaction is framed as
informational: people approach informational political communication
with relatively open evaluations, and AI mediation disrupts this
openness more than it disrupts an already-negative response to
persuasive outreach. The normative penalty for AI involvement may
therefore attach to the mode of delivery rather than the explicit intent
of the communication. This directly challenges the assumption that
nominally nonpartisan or informational AI-enabled civic engagement tools
will be evaluated more charitably than persuasive ones, a challenge with
direct relevance to current proposals for using AI chatbots in
democratically strengthening ways \cite{Velez:2025aa}. If people
penalize AI mediation most when they are otherwise more open to the
interaction, our findings suggest that the campaigns and civic
organizations that stand to lose most from AI deployment are precisely
those conducting the most legitimate-seeming outreach. That said, the AI penalty is not limited to informational outreach: people still evaluated AI-mediated persuasive outreach more negatively than persuasive outreach by a human, and outreach was viewed most negatively when it was both AI-mediated and explicitly persuasive.

Several limitations bear on these conclusions. The design asked
participants to identify a personally important political issue and
reflect on it immediately before treatment --- producing near-ceiling
importance ratings in both countries (US: M = 6.46; UK: M = 6.22). This
was deliberate: most AI persuasion research targets similarly
high-salience topics (e.g., immigration, climate change, conspiracy
beliefs) because these are the conditions under which automated
persuasion is most interesting and most likely to be deployed. Our
design grounds the findings in comparable conditions and makes the
penalties we document directly relevant to the contexts in which AI
persuasion is actively studied. The content analysis of the mentioned
issues further indicates that these contexts were substantively
realistic rather than abstract or idiosyncratic. At the same time,
reactance theory predicts that the magnitude of the persuasion penalty
scales with the importance of the threatened attitude
\cite{Brehm:1981aa,Dillard:2005aa}. The persuasion penalty we
observe represents its magnitude under high-involvement conditions; it
likely overstates what would be found in lower-salience encounters. The
AI penalty, which we interpret as reflecting normative judgment rather
than reactance, is theoretically less sensitive to this feature and may
therefore generalize more broadly across involvement levels.
Additionally, participants evaluated descriptions of anticipated
outreach rather than engaging directly with AI-mediated systems. This
design captures the evaluative responses that precede engagement, but
cannot confirm that these responses translate into behavioral avoidance
under real interaction conditions with actual AI systems. Finally, all conditions specified outreach from a campaign taking a position different from the respondent's own. The penalties we document may therefore be strongest when outreach comes from an opposing side; whether they generalize to contexts where the campaign's position aligns with the respondent's views remains an open question.

Taken together, the findings provide evidence for three results that complicate how
AI-enabled persuasion should be evaluated. First, both persuasive
framing and AI mediation impose consistent evaluative costs on perceived
legitimacy, acceptability, and organizational trust. Those costs remain
hidden when measuring attitude change under enforced exposure and are
not immediately visible in participation rates. Second, these costs are
often most pronounced in the communicative contexts where AI deployment
might otherwise seem most defensible: when outreach is informational, AI
mediation can disrupt evaluations that would otherwise remain open.
Third, the two penalties substitute for rather than compound each other
under conditions of strong initial reactance, a pattern that is itself
sensitive to how normalized persuasive campaign contact is in a given
political environment, a difference that extends to which penalty political orientation most clearly moderates (see Appendix~\ref{section:polor}). This implies that the evaluative consequences of AI-mediated persuasion are not determined solely by the content of the
messages delivered, but also by the broader political communication
context in which deployment occurs. As automated persuasion becomes more
widespread, these evaluative dynamics are likely to shape its
consequences for democratic communication at least as much as its
persuasive effectiveness.

\section*{Acknowledgements}\label{acknowledgements}

The authors used ChatGPT 5.4 and Claude Sonnet 4.6 for language
improvement, editing, and code review. Andreas Jungherr's work was supported by a
grant by the Bavarian State Ministry of Science and the Arts coordinated
by the Bavarian Research Institute for Digital Transformation (bidt).
Adrian Rauchfleisch's work was supported by the National Science and
Technology Council, Taiwan (R.O.C.) (Grant No. 114-2628-H-002-007) and by the Taiwan Social Resilience Research Center
(Grant No.115L9003) from the Higher Education Sprout Project of the
Ministry of Education in Taiwan.

\section*{Author Bios}\label{author-bios}

\noindent\textbf{Andreas Jungherr} holds the Chair for \emph{Political Science,
especially Digital Transformation} at the University of Bamberg and is
Director at the \emph{Bavarian Research Institute for Digital
Transformation (bidt)}. He examines the impact of digital media on
politics and society, with a special focus on Artificial Intelligence,
political communication, and governance. He is the author of
\emph{Retooling Politics: How Digital Media is Shaping Democracy} (with
Gonzalo Rivero and Daniel Gayo-Avello, Cambridge University Press: 2020)
and \emph{Digital Transformations of the Public Arena} (with Ralph
Schroeder, Cambridge University Press: 2022).

\noindent\textbf{Adrian Rauchfleisch}~is a Professor at the Graduate Institute of
Journalism, National Taiwan University. His research focuses on the
interplay of politics, technology, and journalism in Asia, Europe, and
the United States. His new project explores Artificial Intelligence's
influence on society across different cultural contexts.

\bibliographystyle{apacite}
\bibliography{references}

\clearpage
\appendix
\appendix           % this will insert a “Supplementary Information” title page (page,toc,title)
\part{}             % uses the \appendixname, so you get exactly “Supplementary Information”
\parttoc            % prints a mini‐TOC just for this part

\section{Data}
\subsection{Power analysis}
For our preregistration, we conducted a power simulation in R to determine the sample size needed to detect effects similar to those observed in a pretest (Chat/Human = 3.65, Chat/AI = 4.00, Persuasion/Human = 4.10, Persuasion/AI = 4.85/ SD = 1.2). We simulated a 2 (informational vs. persuasive outreach) × 2 (human vs. AI) between-subject design using the Superpower package (1,000 iterations, seed = 2026) in R \citeSI{lakens_simulation-based_2021}. With 450 participants per cell (N = 1,800 total), the simulation yielded 100\% power for both main effects (partial $\eta^{2}$ = .059 for persuasive intent; partial $\eta^{2}$ = .043 for AI vs. human) and 90\% power for the interaction (partial $\eta^{2}$ = .0066).

We also checked with the power simulation two planned simple-effects contrasts (AI vs. human within informational outreach; AI vs. human within persuasive outreach) with adjusted p-values for these contrasts using Holm's method. Based on the simulation, the AI--human contrast is well-powered in the informational condition (98.3\% power; Cohen’s d = 0.268) and in the persuasive condition (100\% power; Cohen’s d = 0.580).

\subsection{Sample and data-quality checks}
\label{section:pop}
We used a number of checks to ensure the quality of our data. First, we used Prolific's new authenticity check that could be integrated into the Qualtrics online survey \citeSI{gordon_authenticity_2026}. In the US, Prolific's tool did not indicate any bots (97.2\% high authenticity, 2.8\% not enough signals or not possible to evaluate). The results of the authenticity check were similar in the UK, with only two users for whom potential bot behavior was indicated (97.8\% high authenticity, 2.1\% not enough signals or not possible to evaluate, 0.1\% with low authenticity). Secondly, we checked with Qualtrics reCAPTCHA, which was activated in our online questionnaires, how many users received a score lower than 0.5, which indicates a participant is likely a bot. In the US, 15 participants, and in the UK, 1 participant were flagged. Lastly, we also included a VPN and proxy check on Qualtrics that relied on iphub.info. In both countries, we only identified a very few users who used a VPN (US=1; UK=3). Also, the indicated location was consistently high for the US (US=1792; other primarily adjacent countries=8) and UK (UK=1745, United States=54; Hungary=1). We checked the UK respondent who accessed the questionnaire from the US, which looked to be UK-focused based on the issues mentioned. Overall, these checks provide little evidence of substantial bot activity, VPN use, or location inconsistencies that would call the quality of the data into question.

\subsubsection{United States}
We recruited participants via the Prolific platform and used quota-based sampling to approximate the US population in terms of age, gender, and party identification. In terms of gender, our sample consisted of 50.11\% women, 48.61\% men, and 1.28\% identifying as another gender. In terms of party identification, 28.50\% of respondents identified as Republican, 31.56\% as Democrat, 38.72\% as Independent, and 1.22\% as other or no preference. We also achieved a good distribution across the different age brackets (see Table \ref{table:age_bracket_distribution_us}).

\begin{table}[H]
\centering
\label{table:age_bracket_distribution_us}
\begin{tabular}{lrr}
\toprule
\textbf{Age Bracket} & \textbf{Count} & \textbf{Percentage (\%)} \\
\midrule
18--27 & 292 & 16.22 \\
28--37 & 337 & 18.72 \\
38--47 & 313 & 17.39 \\
48--57 & 316 & 17.56 \\
58--85 & 542 & 30.11 \\
\bottomrule
\end{tabular}
\caption{Distribution of Sample Across Age Brackets for the US.}
\end{table}

\subsubsection{United Kingdom}
We recruited participants via the Prolific platform and used quota-based sampling to approximate the UK population with respect to age, gender, and party identification. In terms of gender, our sample consisted of 50.61\% women, 49.06\% men, and 0.33\% identifying as another gender. Respondents were also well distributed across the seven party-identification categories. In the UK, 35.94\% of respondents identified with Labour, followed by 18.11\% with the Conservatives, 15.67\% with Reform UK, 13.44\% with the Green Party, 11.00\% with the Liberal Democrats, and 2.67\% with the Scottish National Party (SNP), while 3.17\% selected another party. We also achieved a good distribution across the different age brackets (see Table \ref{table:age_bracket_distribution_uk}).

\begin{table}[H]
\centering
\label{table:age_bracket_distribution_uk}
\begin{tabular}{lrr}
\toprule
\textbf{Age Bracket} & \textbf{Count} & \textbf{Percentage (\%)} \\
\midrule
18--27 & 216 & 12.00 \\
28--37 & 350 & 19.44 \\
38--47 & 312 & 17.33 \\
48--57 & 376 & 20.89 \\
58--85 & 546 & 30.33 \\
\bottomrule
\end{tabular}
\caption{Distribution of Sample Across Age Brackets for the UK.}
\end{table}

\section{Measures}
\label{section:items}
The complete questionnaires are available on OSF:
\href{https://osf.io/zvr7f/overview?view_only=21c8e4c35d3642f79dfa64d1b9d55d87}{Link to the questionnaire} 

\subsection{Descriptive statistics: United States}
\begin{table}[H]
\resizebox{\textwidth}{!}{
\begin{tabular}{L{4.3cm}L{7.2cm}cc}
\toprule
Variable & Question/Operationalization & M (SD) & n\\
\midrule

H1/2a: Willingness to participate 
& If you had the chance to participate in such an exchange, how interested do you think you would be in doing so? (1 = "Not at all interested", 7 = "Very interested") 
& 3.58 (2.05) & 1800\\

H1/2b: Perceived threat to freedom (3 items, $\alpha$ = 0.86) 
& (1 = "Strongly disagree", 7 = "Strongly agree") 
& 3.61 (1.70) & 1800\\
 & The outreach threatened my freedom to choose. 
 & 2.95 (1.82) & 1800\\
 & The outreach tried to make a decision for me. 
 & 3.63 (2.00) & 1800\\
 & The outreach tried to pressure me. 
 & 4.24 (1.98) & 1800\\

H1/2c: Acceptability (3 items, $\alpha$ = 0.79) 
& (1 = "Strongly disagree", 7 = "Strongly agree"; second and third item reverse-coded)   
& 3.63 (1.51) & 1800\\
 & This is an acceptable campaign approach. 
 & 3.84 (1.76) & 1800\\
 & This campaign approach feels manipulative. (-)
 & 4.66 (1.82) & 1800\\
 & This kind of outreach is not how campaigns should operate. (-)
 & 4.29 (1.79) & 1800\\

H1/2d: Positive impact (3 items, $\alpha$ = 0.84) 
& (1 = "Strongly disagree", 7 = "Strongly agree") 
& 3.70 (1.59) & 1800\\
 & This sort of outreach contributes positively to public discourse. 
 & 3.82 (1.76) & 1800\\
 & I appreciate this kind of outreach as an opportunity to learn about political views that differ from my own. 
 & 3.86 (1.87) & 1800\\
 & This sort of outreach helps me make up my mind about the issue at hand. 
 & 3.41 (1.85) & 1800\\

H1/2e: Future campaign avoidance (3 items, $\alpha$ = 0.90) 
& (1 = "Strongly disagree", 7 = "Strongly agree") 
& 4.51 (1.71) & 1800\\
 & If campaigns commonly used this approach, I would try to avoid interacting with them. 
 & 4.65 (1.88) & 1800\\
 & If campaigns commonly used this approach, I would reduce how often I engage with political content. 
 & 4.17 (1.86) & 1800\\
 & If campaigns commonly used this approach, I would be more likely to ignore or block their messages. 
 & 4.71 (1.88) & 1800\\

H1/2f: Penalty for source (4 items, $\alpha$ = 0.84) 
& (1 = "Strongly disagree", 7 = "Strongly agree"; second item reverse-coded) 
& 4.38 (1.36) & 1800\\
 & An organization using this approach is untrustworthy. 
 & 4.12 (1.75) & 1800\\
 & An organization using this approach should be supported. (-) 
 & 3.50 (1.60) & 1800\\
 & Organizations using this approach should be publicly held accountable. 
 & 4.76 (1.61) & 1800\\
 & Organizations that use this approach harm the causes they support. 
 & 4.14 (1.66) & 1800\\

\bottomrule
\end{tabular}
}
\caption{Descriptive statistics for outcome variables and constituent items. (-) indicates negatively formulated items that were recoded for the indices.}
\label{tab:descr_outcomes_us}
\end{table}

\begin{table}[H]
\resizebox{\textwidth}{!}{
\begin{tabular}{L{4.3cm}L{7.2cm}cc}
\toprule
Variable & Question/Operationalization & M (SD) & n\\
\midrule

RQ1: Avoidance of political conversations (4 items, $\alpha$ = 0.88) 
& (1 = "Strongly disagree", 7 = "Strongly agree") 
& 4.39 (1.81) & 1800\\
 & In the past month I have avoided talking politics with family members with whom I disagree. 
 & 4.24 (2.16) & 1800\\
 & In the past month I have avoided talking politics with friends with whom I disagree. 
 & 4.20 (2.08) & 1800\\
 & In the past month I have avoided talking politics with strangers with whom I disagree. 
 & 4.72 (2.12) & 1800\\
 & In the past month I have avoided talking politics with neighbors with whom I disagree. 
 & 4.40 (2.11) & 1800\\

RQ2: Feeling toward people with opposing opinions 
& How would you rate your feelings toward people who support a different position than you on this issue? (0 = "As unfavorable/cold as possible", 100 = "As favorable/warm as possible") 
& 35.71 (29.19) & 1800\\

Feeling toward people with same opinion 
& How would you rate your feelings toward people who support the same position as you on this issue? (0 = "As unfavorable/cold as possible", 100 = "As favorable/warm as possible") 
& 82.58 (16.82) & 1800\\

Affective polarization 
& (in-group warmth - out-group warmth) 
& 46.87 (35.71) & 1800\\

RQ3: AI risk perception (4 items, $\alpha$ = 0.82) 
& (1 = "Strongly disagree", 7 = "Strongly agree") 
& 5.15 (1.32) & 1800\\
 & AI is likely to cause widespread job displacement and unemployment. 
 & 5.10 (1.60) & 1800\\
 & As AI increasingly takes over decision-making, we risk losing control over our lives. 
 & 4.83 (1.75) & 1800\\
 & AI in military applications can lead to unintended escalations of conflicts due to lack of human judgement. 
 & 5.21 (1.61) & 1800\\
 & Unchecked AI development could pose existential threats to humanity. 
 & 5.48 (1.61) & 1800\\

Issue importance 
& How important is this issue to you? (1 = "Not at all important", 7 = "Extremely important") 
& 6.46 (0.79) & 1800\\

Political orientation 
& (1 = "Left", 7 = "Right") 
& 3.73 (1.89) & 1800\\

Age 
& (in years) 
& 46.22 (16.58) & 1800\\

Gender 
& (1 = male) 
& 48.6\% & 1800\\

Education 
& (1 = post-graduate degree / Master degree or higher) 
& 15.8\% & 1800\\

\midrule
Treatment group sizes & Informational human-outreach &  & 443\\
 & Informational AI-outreach &  & 451\\
 & Persuasive human-outreach &  & 452\\
 & Persuasive AI-outreach &  & 454\\

\bottomrule
\end{tabular}
}
\caption{Descriptive statistics for moderators, additional variables, and demographics.}
\label{tab:descr_first}
\end{table}

\subsection{Descriptive statistics: United Kingdom}

\begin{table}[H]
\resizebox{\textwidth}{!}{
\begin{tabular}{L{4.3cm}L{7.2cm}cc}
\toprule
Variable & Question/Operationalization & M (SD) & n\\
\midrule

H1/2a: Willingness to participate 
& If you had the chance to participate in such an exchange, how interested do you think you would be in doing so? (1 = "Not at all interested", 7 = "Very interested") 
& 3.80 (1.90) & 1800\\
H1/2b: Perceived threat to freedom (3 items, $\alpha$ = 0.87) 
& (1 = "Strongly disagree", 7 = "Strongly agree") 
& 3.81 (1.59) & 1800\\
 & The outreach threatened my freedom to choose. 
 & 3.23 (1.73) & 1800\\
 & The outreach tried to make a decision for me. 
 & 3.88 (1.81) & 1800\\
 & The outreach tried to pressure me. 
 & 4.32 (1.81) & 1800\\

H1/2c: Acceptability (3 items, $\alpha$ = 0.80) 
& (1 = "Strongly disagree", 7 = "Strongly agree"; second and third item reverse-coded) 
& 3.59 (1.38) & 1800\\
 & This is an acceptable campaign approach. 
 & 3.86 (1.62) & 1800\\
 & This campaign approach feels manipulative. (-)
 & 4.69 (1.67) & 1800\\
 & This kind of outreach is not how campaigns should operate. (-)
 & 4.42 (1.63) & 1800\\

H1/2d: Positive impact (3 items, $\alpha$ = 0.84) 
& (1 = "Strongly disagree", 7 = "Strongly agree") 
& 3.87 (1.42) & 1800\\
 & This sort of outreach contributes positively to public discourse. 
 & 3.94 (1.56) & 1800\\
 & I appreciate this kind of outreach as an opportunity to learn about political views that differ from my own. 
 & 4.07 (1.66) & 1800\\
 & This sort of outreach helps me make up my mind about the issue at hand. 
 & 3.59 (1.69) & 1800\\

H1/2e: Future campaign avoidance (3 items, $\alpha$ = 0.90) 
& (1 = "Strongly disagree", 7 = "Strongly agree") 
& 4.50 (1.59) & 1800\\
 & If campaigns commonly used this approach, I would try to avoid interacting with them. 
 & 4.62 (1.73) & 1800\\
 & If campaigns commonly used this approach, I would reduce how often I engage with political content. 
 & 4.19 (1.73) & 1800\\
 & If campaigns commonly used this approach, I would be more likely to ignore or block their messages. 
 & 4.68 (1.76) & 1800\\

H1/2f: Penalty for source (4 items, $\alpha$ = 0.84) 
& (1 = "Strongly disagree", 7 = "Strongly agree"; second item reverse-coded) 
& 4.53 (1.24) & 1800\\
 & An organization using this approach is untrustworthy. 
 & 4.19 (1.58) & 1800\\
 & An organization using this approach should be supported. (-) 
 & 3.40 (1.44) & 1800\\
 & Organizations using this approach should be publicly held accountable. 
 & 5.08 (1.47) & 1800\\
 & Organizations that use this approach harm the causes they support. 
 & 4.27 (1.50) & 1800\\

\bottomrule
\end{tabular}
}
\caption{Descriptive statistics for outcome variables and constituent items. (-) indicates negatively formulated items that were recoded for the indices.}
\label{tab:descr_outcomes_uk}
\end{table}

\begin{table}[H]
\resizebox{\textwidth}{!}{
\begin{tabular}{L{4.3cm}L{7.2cm}cc}
\toprule
Variable & Question/Operationalization & M (SD) & n\\
\midrule

RQ1: Avoidance of political conversations (4 items, $\alpha$ = 0.86) 
& (1 = "Strongly disagree", 7 = "Strongly agree") 
& 3.84 (1.66) & 1800\\
 & In the past month I have avoided talking politics with family members with whom I disagree. 
 & 3.60 (1.97) & 1800\\
 & In the past month I have avoided talking politics with friends with whom I disagree. 
 & 3.67 (1.92) & 1800\\
 & In the past month I have avoided talking politics with strangers with whom I disagree. 
 & 4.24 (2.03) & 1800\\
 & In the past month I have avoided talking politics with neighbors with whom I disagree. 
 & 3.83 (1.98) & 1800\\

RQ2: Feeling toward people with opposing opinions 
& How would you rate your feelings toward people who support a different position than you on this issue?  (0 = "As unfavorable/cold as possible", 100 = "As favorable/warm as possible") 
& 36.55 (25.24) & 1800\\

Feeling toward people with same opinion 
& How would you rate your feelings toward people who support the same position as you on this issue?  (0 = "As unfavorable/cold as possible", 100 = "As favorable/warm as possible") 
& 79.38 (16.55) & 1800\\

Affective polarization 
& (in-group warmth - out-group warmth) 
& 42.82 (32.70) & 1800\\

RQ3: AI risk perception (4 items, $\alpha$ = 0.81) 
& (1 = "Strongly disagree", 7 = "Strongly agree") 
& 5.07 (1.16) & 1800\\
 & AI is likely to cause widespread job displacement and unemployment. 
 & 5.07 (1.41) & 1800\\
 & As AI increasingly takes over decision-making, we risk losing control over our lives. 
 & 4.82 (1.57) & 1800\\
 & AI in military applications can lead to unintended escalations of conflicts due to lack of human judgement. 
 & 5.00 (1.41) & 1800\\
 & Unchecked AI development could pose existential threats to humanity. 
 & 5.40 (1.44) & 1800\\

Issue importance 
& How important is this issue to you?  (1 = "Not at all important", 7 = "Extremely important") 
& 6.22 (0.90) & 1800\\

Political orientation 
& (1 = "Left", 7 = "Right") 
& 3.66 (1.40) & 1800\\

Age 
& (in years) 
& 47.45 (15.69) & 1800\\

Gender 
& (1 = male) 
& 49.1\% & 1800\\

Education 
& (1 = post-graduate degree / Master degree or higher) 
& 19.9\% & 1800\\

\midrule
Treatment group sizes & Informational human-outreach &  & 446\\
 & Informational AI-outreach &  & 449\\
 & Persuasive human-outreach &  & 452\\
 & Persuasive AI-outreach &  & 453\\

\bottomrule
\end{tabular}
}
\caption{Descriptive statistics for moderators, additional variables, and demographics.}
\label{tab:descr_moderators}
\end{table}

\subsection{Treatments}
\label{section:treatments}
In this section, we present the treatments we have used for the experiment. Before people were shown the treatment text, we used the following text, preparing them for the following page: "On the next page, you will see an example of outreach from a campaign about the issue you just named. The campaign supports a different position than yours. Please read the message carefully. We will then ask a few questions. The “Continue” button will appear after 10 seconds."

\subsubsection{Informational Human-outreach}
Imagine you are contacted by a campaigner to talk about the issue you just named. The campaign takes a position that is different from yours. The campaigner shares information and answers your questions on the campaign's positions, leaving it to you what to do with this information.

\subsubsection{Informational AI-outreach}
Imagine you are contacted by an AI campaign chatbot to talk about the issue you just named. The campaign takes a position that is different from yours. The AI chatbot shares information and answers your questions on the campaign's positions, leaving it to you what to do with this information.

\subsubsection{Persuasive human-outreach}
Imagine you are contacted by a campaigner to talk about the issue you just named. The campaign takes a position that is different from yours. The campaigner tries to change your opinion in line with the campaign’s goals by using persuasion techniques to make messages more convincing. The campaigner adapts messages and arguments to what you say about your views and concerns during the conversation to better change your opinion.

\subsubsection{Persuasive AI-outreach}
Imagine you are contacted by an AI campaign chatbot to talk about the issue you just named. The campaign takes a position that is different from yours. The AI chatbot tries to change your opinion in line with the campaign’s goals by using persuasion techniques to make messages more convincing. The AI chatbot adapts messages and arguments to what you say about your views and concerns during the conversation to better change your opinion.

\section{Model results}
\label{section:models}
We tested hypotheses using OLS regression models with effect coding for both factors (-0.5, +0.5) and their interaction. With this coding scheme, the coefficients for the main effects represent the mean difference between the two levels of each factor, averaged over the other factor.
\subsection{Main hypotheses}
\subsubsection{United States}
\begin{table}[H]
\centering
\begin{tabular}[t]{lccc}
\toprule
Predictors & Estimate & 95\% CI & p \\
\midrule
Intercept & 3.58 & [3.49, 3.68] & <0.001 \\
H1a: Persuasive outreach & -0.18 & [-0.37, 0.01] & 0.059 \\
H2a: AI-mediated outreach & 0.02 & [-0.17, 0.21] & 0.820 \\
H3: Persuasive $\times$ AI-mediated outreach & 0.25 & [-0.13, 0.63] & 0.202 \\
\addlinespace
Observations & 1800 &  &  \\
$R^2$ / adjusted $R^2$ & 0.003 / 0.001 &  &  \\
\bottomrule
\end{tabular}
\caption{OLS regression model for willingness to participate with 95\% confidence intervals. $N = 1800$.}
\end{table}

\begin{table}[H]
\centering
\begin{tabular}[t]{lccc}
\toprule
Predictors & Estimate & 95\% CI & p \\
\midrule
Intercept & 3.60 & [3.53, 3.68] & <0.001 \\
H1b: Persuasive outreach & 0.87 & [0.72, 1.02] & <0.001 \\
H2b: AI-mediated outreach & 0.19 & [0.04, 0.34] & 0.016 \\
H3: Persuasive $\times$ AI-mediated outreach & -0.17 & [-0.48, 0.13] & 0.269 \\
\addlinespace
Observations & 1800 &  &  \\
$R^2$ / adjusted $R^2$ & 0.068 / 0.067 &  &  \\
\bottomrule
\end{tabular}
\caption{OLS regression model for perceived threat to freedom with 95\% confidence intervals. $N = 1800$.}
\end{table}

\begin{table}[H]
\centering
\begin{tabular}[t]{lccc}
\toprule
Predictors & Estimate & 95\% CI & p \\
\midrule
Intercept & 3.63 & [3.57, 3.70] & <0.001 \\
H1c: Persuasive outreach & -0.50 & [-0.63, -0.37] & <0.001 \\
H2c: AI-mediated outreach & -0.58 & [-0.72, -0.45] & <0.001 \\
H3: Persuasive $\times$ AI-mediated outreach & 0.20 & [-0.07, 0.47] & 0.148 \\
\addlinespace
Observations & 1800 &  &  \\
$R^2$ / adjusted $R^2$ & 0.066 / 0.064 &  &  \\
\bottomrule
\end{tabular}
\caption{OLS regression model for acceptability with 95\% confidence intervals. $N = 1800$.}
\end{table}

\begin{table}[H]
\centering
\begin{tabular}[t]{lccc}
\toprule
Predictors & Estimate & 95\% CI & p \\
\midrule
Intercept & 3.70 & [3.63, 3.77] & <0.001 \\
H1d: Persuasive outreach & -0.30 & [-0.44, -0.15] & <0.001 \\
H2d: AI-mediated outreach & -0.44 & [-0.58, -0.29] & <0.001 \\
H3: Persuasive $\times$ AI-mediated outreach & 0.08 & [-0.21, 0.37] & 0.600 \\
\addlinespace
Observations & 1800 &  &  \\
$R^2$ / adjusted $R^2$ & 0.028 / 0.026 &  &  \\
\bottomrule
\end{tabular}
\caption{OLS regression model for perceived positive impact of campaign outreach with 95\% confidence intervals. $N = 1800$.}
\end{table}

\begin{table}[H]
\centering
\begin{tabular}[t]{lccc}
\toprule
Predictors & Estimate & 95\% CI & p \\
\midrule
Intercept & 4.51 & [4.43, 4.59] & <0.001 \\
H1e: Persuasive outreach & 0.42 & [0.26, 0.58] & <0.001 \\
H2e: AI-mediated outreach & 0.38 & [0.23, 0.54] & <0.001 \\
H3: Persuasive $\times$ AI-mediated outreach & -0.27 & [-0.58, 0.04] & 0.093 \\
\addlinespace
Observations & 1800 &  &  \\
$R^2$ / adjusted $R^2$ & 0.029 / 0.027 &  &  \\
\bottomrule
\end{tabular}
\caption{OLS regression model for future campaign avoidance with 95\% confidence intervals. $N = 1800$.}
\end{table}

\begin{table}[H]
\centering
\begin{tabular}[t]{lccc}
\toprule
Predictors & Estimate & 95\% CI & p \\
\midrule
Intercept & 4.38 & [4.32, 4.44] & <0.001 \\
H1f: Persuasive outreach & 0.38 & [0.26, 0.50] & <0.001 \\
H2f: AI-mediated outreach & 0.55 & [0.43, 0.68] & <0.001 \\
H3: Persuasive $\times$ AI-mediated outreach & -0.13 & [-0.38, 0.11] & 0.293 \\
\addlinespace
Observations & 1800 &  &  \\
$R^2$ / adjusted $R^2$ & 0.061 / 0.060 &  &  \\
\bottomrule
\end{tabular}
\caption{OLS regression model for penalty for source with 95\% confidence intervals. $N = 1800$.}
\end{table}

\subsubsection{United Kingdom}
\begin{table}[H]
\centering
\begin{tabular}[t]{lccc}
\toprule
Predictors & Estimate & 95\% CI & p \\
\midrule
Intercept & 3.80 & [3.71, 3.89] & <0.001 \\
H1a: Persuasive outreach & -0.23 & [-0.41, -0.06] & 0.009 \\
H2a: AI-mediated outreach & -0.03 & [-0.21, 0.14] & 0.696 \\
H3: Persuasive $\times$ AI-mediated outreach & 0.40 & [0.05, 0.75] & 0.025 \\
\addlinespace
Observations & 1800 &  &  \\
$R^2$ / adjusted $R^2$ & 0.007 / 0.005 &  &  \\
\bottomrule
\end{tabular}
\caption{OLS regression model for willingness to participate with 95\% confidence intervals. $N = 1800$.}
\end{table}

\begin{table}[H]
\centering
\begin{tabular}[t]{lccc}
\toprule
Predictors & Estimate & 95\% CI & p \\
\midrule
Intercept & 3.81 & [3.73, 3.88] & <0.001 \\
H1b: Persuasive outreach & 0.79 & [0.64, 0.93] & <0.001 \\
H2b: AI-mediated outreach & 0.27 & [0.13, 0.42] & <0.001 \\
H3: Persuasive $\times$ AI-mediated outreach & -0.25 & [-0.54, 0.03] & 0.083 \\
\addlinespace
Observations & 1800 &  &  \\
$R^2$ / adjusted $R^2$ & 0.070 / 0.068 &  &  \\
\bottomrule
\end{tabular}
\caption{OLS regression model for perceived threat to freedom with 95\% confidence intervals. $N = 1800$.}
\end{table}

\begin{table}[H]
\centering
\begin{tabular}[t]{lccc}
\toprule
Predictors & Estimate & 95\% CI & p \\
\midrule
Intercept & 3.59 & [3.53, 3.65] & <0.001 \\
H1c: Persuasive outreach & -0.57 & [-0.69, -0.44] & <0.001 \\
H2c: AI-mediated outreach & -0.55 & [-0.67, -0.43] & <0.001 \\
H3: Persuasive $\times$ AI-mediated outreach & 0.21 & [-0.04, 0.45] & 0.101 \\
\addlinespace
Observations & 1800 &  &  \\
$R^2$ / adjusted $R^2$ & 0.083 / 0.081 &  &  \\
\bottomrule
\end{tabular}
\caption{OLS regression model for acceptability with 95\% confidence intervals. $N = 1800$.}
\end{table}

\begin{table}[H]
\centering
\begin{tabular}[t]{lccc}
\toprule
Predictors & Estimate & 95\% CI & p \\
\midrule
Intercept & 3.87 & [3.81, 3.94] & <0.001 \\
H1d: Persuasive outreach & -0.31 & [-0.44, -0.18] & <0.001 \\
H2d: AI-mediated outreach & -0.31 & [-0.44, -0.18] & <0.001 \\
H3: Persuasive $\times$ AI-mediated outreach & 0.39 & [0.13, 0.65] & 0.003 \\
\addlinespace
Observations & 1800 &  &  \\
$R^2$ / adjusted $R^2$ & 0.028 / 0.027 &  &  \\
\bottomrule
\end{tabular}
\caption{OLS regression model for perceived positive impact of campaign outreach with 95\% confidence intervals. $N = 1800$.}
\end{table}

\begin{table}[H]
\centering
\begin{tabular}[t]{lccc}
\toprule
Predictors & Estimate & 95\% CI & p \\
\midrule
Intercept & 4.49 & [4.42, 4.57] & <0.001 \\
H1e: Persuasive outreach & 0.56 & [0.42, 0.70] & <0.001 \\
H2e: AI-mediated outreach & 0.50 & [0.35, 0.64] & <0.001 \\
H3: Persuasive $\times$ AI-mediated outreach & -0.35 & [-0.64, -0.06] & 0.017 \\
\addlinespace
Observations & 1800 &  &  \\
$R^2$ / adjusted $R^2$ & 0.058 / 0.056 &  &  \\
\bottomrule
\end{tabular}
\caption{OLS regression model for future campaign avoidance with 95\% confidence intervals. $N = 1800$.}
\end{table}

\begin{table}[H]
\centering
\begin{tabular}[t]{lccc}
\toprule
Predictors & Estimate & 95\% CI & p \\
\midrule
Intercept & 4.53 & [4.48, 4.59] & <0.001 \\
H1f: Persuasive outreach & 0.40 & [0.29, 0.51] & <0.001 \\
H2f: AI-mediated outreach & 0.48 & [0.37, 0.59] & <0.001 \\
H3: Persuasive $\times$ AI-mediated outreach & -0.30 & [-0.52, -0.08] & 0.008 \\
\addlinespace
Observations & 1800 &  &  \\
$R^2$ / adjusted $R^2$ & 0.068 / 0.066 &  &  \\
\bottomrule
\end{tabular}
\caption{OLS regression model for penalty for source with 95\% confidence intervals. $N = 1800$.}
\end{table}

\subsubsection{Effect size}

\begin{table}[H]
\centering
\small
\begin{tabular}{L{4.8cm}L{4.0cm}L{4.0cm}}
\toprule
Outcome & Persuasion penalty & AI penalty \\
\midrule
H1a/H2a: Willingness to participate
& -0.09 [-0.18, 0.00]
& 0.01 [-0.08, 0.10] \\

H1b/H2b: Perceived threat to freedom
& 0.53 [0.43, 0.62]
& 0.11 [0.02, 0.20] \\

H1c/H2c: Acceptability
& -0.33 [-0.43, -0.24]
& -0.39 [-0.49, -0.30] \\

H1d/H2d: Positive impact
& -0.19 [-0.28, -0.10]
& -0.28 [-0.37, -0.18] \\

H1e/H2e: Future campaign avoidance
& 0.25 [0.15, 0.34]
& 0.22 [0.13, 0.32] \\

H1f/H2f: Source evaluation
& 0.28 [0.19, 0.37]
& 0.41 [0.32, 0.51] \\
\bottomrule
\end{tabular}
\caption{Cohen's $d$ effect sizes for the two main predictors across all preregistered outcomes in the United States. Entries are Cohen's $d$ with 95\% confidence intervals, signed to match the reporting direction used in the main analyses.}
\label{tab:cohens_d_us}
\end{table}

\begin{table}[H]
\centering
\small
\begin{tabular}{L{4.8cm}L{4.0cm}L{4.0cm}}
\toprule
Outcome & Persuasion penalty & AI penalty \\
\midrule
H1a/H2a: Willingness to participate
& -0.12 [-0.22, -0.03]
& -0.02 [-0.11, 0.07] \\

H1b/H2b: Perceived threat to freedom
& 0.51 [0.42, 0.60]
& 0.17 [0.08, 0.26] \\

H1c/H2c: Acceptability
& -0.42 [-0.51, -0.32]
& -0.41 [-0.50, -0.31] \\

H1d/H2d: Positive impact
& -0.22 [-0.31, -0.12]
& -0.22 [-0.31, -0.13] \\

H1e/H2e: Future campaign avoidance
& 0.36 [0.26, 0.45]
& 0.31 [0.22, 0.41] \\

H1f/H2f: Source evaluation
& 0.33 [0.23, 0.42]
& 0.40 [0.30, 0.49] \\
\bottomrule
\end{tabular}
\caption{Cohen's $d$ effect sizes for the two main predictors across all preregistered outcomes in the United Kingdom. Entries are Cohen's $d$ with 95\% confidence intervals, signed to match the reporting direction used in the main analyses.}
\label{tab:cohens_d_uk}
\end{table}

\subsection{Preregistered planned contrasts}
\label{section:contrasts}
Following preregistration, we conducted planned simple effect contrasts to compare AI- and human-mediated outreach separately within informational and persuasive conditions. The P-values for these contrasts were adjusted using Holm's method.

\subsubsection{United States}
\begin{table}[H]
\centering
\begin{tabular}{L{3.6cm}L{5.2cm}L{5.2cm}}
\toprule
Outcome & Information outreach: Human $-$ AI & Persuasive outreach: Human $-$ AI \\
\midrule
H1a/H2a: Willingness to participate 
& 0.10 [-0.17, 0.37], $p_{\mathrm{Holm}} = .460$
& -0.15 [-0.41, 0.12], $p_{\mathrm{Holm}} = .286$ \\

H1b/H2b: Perceived threat to freedom 
& -0.27 [-0.49, -0.06], $p_{\mathrm{Holm}} = .013$
& -0.10 [-0.32, 0.11], $p_{\mathrm{Holm}} = .350$ \\

H1c/H2c: Acceptability of campaign outreach 
& 0.68 [0.49, 0.87], $p_{\mathrm{Holm}} < .001$
& 0.48 [0.29, 0.67], $p_{\mathrm{Holm}} < .001$ \\

H1d/H2d: Perceived positive impact of campaign outreach 
& 0.48 [0.27, 0.68], $p_{\mathrm{Holm}} < .001$
& 0.40 [0.20, 0.61], $p_{\mathrm{Holm}} < .001$ \\

H1e/H2e: Future campaign avoidance 
& -0.51 [-0.74, -0.29], $p_{\mathrm{Holm}} < .001$
& -0.25 [-0.47, -0.03], $p_{\mathrm{Holm}} = .027$ \\

H1f/H2f: Penalty for source 
& -0.62 [-0.79, -0.45], $p_{\mathrm{Holm}} < .001$
& -0.49 [-0.66, -0.32], $p_{\mathrm{Holm}} < .001$ \\
\bottomrule
\end{tabular}
\caption{Preregistered planned contrasts comparing human-mediated and AI-mediated outreach within each outreach condition. Entries are estimated mean differences (Human $-$ AI) with 95\% confidence intervals and Holm-adjusted p-values. Positive values indicate higher outcome values for human-mediated outreach; negative values indicate higher outcome values for AI-mediated outreach. $N = 1800$.}
\end{table}

\subsubsection{United Kingdom}

\begin{table}[H]
\centering
\begin{tabular}{L{3.6cm}L{5.2cm}L{5.2cm}}
\toprule
Outcome & Information outreach: Human $-$ AI & Persuasive outreach: Human $-$ AI \\
\midrule
H1a/H2a: Willingness to participate 
& 0.24 [-0.01, 0.49], $p_{\mathrm{Holm}} = .063$
& -0.17 [-0.41, 0.08], $p_{\mathrm{Holm}} = .188$ \\

H1b/H2b: Perceived threat to freedom 
& -0.40 [-0.60, -0.20], $p_{\mathrm{Holm}} < .001$
& -0.15 [-0.35, 0.05], $p_{\mathrm{Holm}} = .151$ \\

H1c/H2c: Acceptability of campaign outreach 
& 0.65 [0.48, 0.83], $p_{\mathrm{Holm}} < .001$
& 0.45 [0.28, 0.62], $p_{\mathrm{Holm}} < .001$ \\

H1d/H2d: Perceived positive impact of campaign outreach 
& 0.51 [0.32, 0.69], $p_{\mathrm{Holm}} < .001$
& 0.12 [-0.07, 0.30], $p_{\mathrm{Holm}} = .210$ \\

H1e/H2e: Future campaign avoidance 
& -0.67 [-0.87, -0.47], $p_{\mathrm{Holm}} < .001$
& -0.32 [-0.52, -0.12], $p_{\mathrm{Holm}} = .002$ \\

H1f/H2f: Penalty for source 
& -0.63 [-0.79, -0.48], $p_{\mathrm{Holm}} < .001$
& -0.33 [-0.49, -0.18], $p_{\mathrm{Holm}} < .001$ \\
\bottomrule
\end{tabular}
\caption{Preregistered planned contrasts comparing human-mediated and AI-mediated outreach within each outreach condition. Entries are estimated mean differences (Human $-$ AI) with 95\% confidence intervals and Holm-adjusted p-values. Positive values indicate higher outcome values for human-mediated outreach; negative values indicate higher outcome values for AI-mediated outreach. $N = 1800$.}
\end{table}

\subsection{Moderators}
\label{section:moderators}
We also preregistered three moderator variables. In this section, we report the interactions between these three moderators and the human vs. AI outreach variable.

\subsubsection{Avoidance of political conversations}
\medskip \noindent\textbf{United States}

\begin{table}[H]
\centering
\begin{tabular}{L{5.8cm}ccc}
\toprule
Predictors & Estimate & 95\% CI & p \\
\midrule
Intercept & 3.58 & [3.49, 3.68] & <0.001 \\
Persuasive outreach & -0.19 & [-0.38, -0.00] & 0.048 \\
AI-mediated outreach & 0.02 & [-0.17, 0.21] & 0.827 \\
Avoidance of political conversations & -0.12 & [-0.17, -0.07] & <0.001 \\
Persuasive $\times$ AI-mediated outreach & 0.24 & [-0.13, 0.62] & 0.204 \\
RQ1: AI-mediated outreach $\times$ Avoidance of political conversations & 0.10 & [-0.01, 0.20] & 0.070 \\
\addlinespace
Observations & 1800 &  &  \\
$R^2$ / adjusted $R^2$ & 0.015 / 0.013 &  &  \\
\bottomrule
\end{tabular}
\caption{OLS regression model for willingness to participate with 95\% confidence intervals, including avoidance of political conversations as a moderator. $N = 1800$.}
\end{table}

\begin{table}[H]
\centering
\begin{tabular}{L{5.8cm}ccc}
\toprule
Predictors & Estimate & 95\% CI & p \\
\midrule
Intercept & 3.60 & [3.53, 3.68] & <0.001 \\
Persuasive outreach & 0.88 & [0.73, 1.03] & <0.001 \\
AI-mediated outreach & 0.19 & [0.04, 0.34] & 0.014 \\
Avoidance of political conversations & 0.11 & [0.07, 0.15] & <0.001 \\
Persuasive $\times$ AI-mediated outreach & -0.16 & [-0.46, 0.14] & 0.297 \\
RQ1: AI-mediated outreach $\times$ Avoidance of political conversations & 0.02 & [-0.06, 0.10] & 0.655 \\
\addlinespace
Observations & 1800 &  &  \\
$R^2$ / adjusted $R^2$ & 0.081 / 0.079 &  &  \\
\bottomrule
\end{tabular}
\caption{OLS regression model for perceived threat to freedom with 95\% confidence intervals, including avoidance of political conversations as a moderator. $N = 1800$.}
\end{table}

\begin{table}[H]
\centering
\begin{tabular}{L{5.8cm}ccc}
\toprule
Predictors & Estimate & 95\% CI & p \\
\midrule
Intercept & 3.63 & [3.57, 3.70] & <0.001 \\
Persuasive outreach & -0.50 & [-0.64, -0.37] & <0.001 \\
AI-mediated outreach & -0.58 & [-0.72, -0.45] & <0.001 \\
Avoidance of political conversations & -0.05 & [-0.09, -0.01] & 0.008 \\
Persuasive $\times$ AI-mediated outreach & 0.20 & [-0.07, 0.46] & 0.155 \\
RQ1: AI-mediated outreach $\times$ Avoidance of political conversations & 0.01 & [-0.06, 0.09] & 0.777 \\
\addlinespace
Observations & 1800 &  &  \\
$R^2$ / adjusted $R^2$ & 0.069 / 0.067 &  &  \\
\bottomrule
\end{tabular}
\caption{OLS regression model for acceptability of campaign outreach with 95\% confidence intervals, including avoidance of political conversations as a moderator. $N = 1800$.}
\end{table}

\begin{table}[H]
\centering
\begin{tabular}{L{5.8cm}ccc}
\toprule
Predictors & Estimate & 95\% CI & p \\
\midrule
Intercept & 3.70 & [3.63, 3.77] & <0.001 \\
Persuasive outreach & -0.30 & [-0.45, -0.16] & <0.001 \\
AI-mediated outreach & -0.44 & [-0.58, -0.29] & <0.001 \\
Avoidance of political conversations & -0.04 & [-0.08, -0.00] & 0.036 \\
Persuasive $\times$ AI-mediated outreach & 0.08 & [-0.21, 0.37] & 0.590 \\
RQ1: AI-mediated outreach $\times$ Avoidance of political conversations & 0.07 & [-0.01, 0.15] & 0.085 \\
\addlinespace
Observations & 1800 &  &  \\
$R^2$ / adjusted $R^2$ & 0.032 / 0.029 &  &  \\
\bottomrule
\end{tabular}
\caption{OLS regression model for perceived positive impact of campaign outreach with 95\% confidence intervals, including avoidance of political conversations as a moderator. $N = 1800$.}
\end{table}

\begin{table}[H]
\centering
\begin{tabular}{L{5.8cm}ccc}
\toprule
Predictors & Estimate & 95\% CI & p \\
\midrule
Intercept & 4.51 & [4.43, 4.59] & <0.001 \\
Persuasive outreach & 0.43 & [0.28, 0.59] & <0.001 \\
AI-mediated outreach & 0.38 & [0.23, 0.54] & <0.001 \\
Avoidance of political conversations & 0.15 & [0.11, 0.19] & <0.001 \\
Persuasive $\times$ AI-mediated outreach & -0.25 & [-0.56, 0.05] & 0.106 \\
RQ1: AI-mediated outreach $\times$ Avoidance of political conversations & 0.00 & [-0.08, 0.09] & 0.991 \\
\addlinespace
Observations & 1800 &  &  \\
$R^2$ / adjusted $R^2$ & 0.054 / 0.052 &  &  \\
\bottomrule
\end{tabular}
\caption{OLS regression model for future campaign avoidance with 95\% confidence intervals, including avoidance of political conversations as a moderator. $N = 1800$.}
\end{table}

\begin{table}[H]
\centering
\begin{tabular}{L{5.8cm}ccc}
\toprule
Predictors & Estimate & 95\% CI & p \\
\midrule
Intercept & 4.38 & [4.32, 4.44] & <0.001 \\
Persuasive outreach & 0.38 & [0.26, 0.50] & <0.001 \\
AI-mediated outreach & 0.55 & [0.43, 0.68] & <0.001 \\
Avoidance of political conversations & 0.05 & [0.02, 0.09] & 0.001 \\
Persuasive $\times$ AI-mediated outreach & -0.13 & [-0.37, 0.12] & 0.305 \\
RQ1: AI-mediated outreach $\times$ Avoidance of political conversations & -0.02 & [-0.08, 0.05] & 0.634 \\
\addlinespace
Observations & 1800 &  &  \\
$R^2$ / adjusted $R^2$ & 0.067 / 0.064 &  &  \\
\bottomrule
\end{tabular}
\caption{OLS regression model for penalty for source with 95\% confidence intervals, including avoidance of political conversations as a moderator. $N = 1800$.}
\end{table}

\medskip \noindent\textbf{United Kingdom}
\begin{table}[H]
\centering
\begin{tabular}{L{5.8cm}ccc}
\toprule
Predictors & Estimate & 95\% CI & p \\
\midrule
Intercept & 3.80 & [3.71, 3.89] & <0.001 \\
Persuasive outreach & -0.23 & [-0.41, -0.06] & 0.009 \\
AI-mediated outreach & -0.03 & [-0.21, 0.14] & 0.717 \\
Avoidance of political conversations & -0.10 & [-0.16, -0.05] & <0.001 \\
Persuasive $\times$ AI-mediated outreach & 0.41 & [0.06, 0.76] & 0.021 \\
RQ1: AI-mediated outreach $\times$ Avoidance of political conversations & 0.15 & [0.05, 0.26] & 0.004 \\
\addlinespace
Observations & 1800 &  &  \\
$R^2$ / adjusted $R^2$ & 0.020 / 0.017 &  &  \\
\bottomrule
\end{tabular}
\caption{OLS regression model for willingness to participate with 95\% confidence intervals, including avoidance of political conversations as a moderator. $N = 1800$.}
\end{table}

\begin{table}[H]
\centering
\begin{tabular}{L{5.8cm}ccc}
\toprule
Predictors & Estimate & 95\% CI & p \\
\midrule
Intercept & 3.81 & [3.74, 3.88] & <0.001 \\
Persuasive outreach & 0.78 & [0.64, 0.92] & <0.001 \\
AI-mediated outreach & 0.27 & [0.13, 0.41] & <0.001 \\
Avoidance of political conversations & 0.15 & [0.10, 0.19] & <0.001 \\
Persuasive $\times$ AI-mediated outreach & -0.27 & [-0.55, 0.01] & 0.057 \\
RQ1: AI-mediated outreach $\times$ Avoidance of political conversations & -0.09 & [-0.18, -0.01] & 0.034 \\
\addlinespace
Observations & 1800 &  &  \\
$R^2$ / adjusted $R^2$ & 0.096 / 0.093 &  &  \\
\bottomrule
\end{tabular}
\caption{OLS regression model for perceived threat to freedom with 95\% confidence intervals, including avoidance of political conversations as a moderator. $N = 1800$.}
\end{table}

\begin{table}[H]
\centering
\begin{tabular}{L{5.8cm}ccc}
\toprule
Predictors & Estimate & 95\% CI & p \\
\midrule
Intercept & 3.59 & [3.53, 3.65] & <0.001 \\
Persuasive outreach & -0.56 & [-0.68, -0.44] & <0.001 \\
AI-mediated outreach & -0.55 & [-0.67, -0.43] & <0.001 \\
Avoidance of political conversations & -0.09 & [-0.12, -0.05] & <0.001 \\
Persuasive $\times$ AI-mediated outreach & 0.22 & [-0.02, 0.46] & 0.078 \\
RQ1: AI-mediated outreach $\times$ Avoidance of political conversations & 0.03 & [-0.05, 0.10] & 0.449 \\
\addlinespace
Observations & 1800 &  &  \\
$R^2$ / adjusted $R^2$ & 0.094 / 0.092 &  &  \\
\bottomrule
\end{tabular}
\caption{OLS regression model for acceptability of campaign outreach with 95\% confidence intervals, including avoidance of political conversations as a moderator. $N = 1800$.}
\end{table}

\begin{table}[H]
\centering
\begin{tabular}{L{5.8cm}ccc}
\toprule
Predictors & Estimate & 95\% CI & p \\
\midrule
Intercept & 3.87 & [3.81, 3.94] & <0.001 \\
Persuasive outreach & -0.31 & [-0.44, -0.18] & <0.001 \\
AI-mediated outreach & -0.31 & [-0.44, -0.18] & <0.001 \\
Avoidance of political conversations & -0.05 & [-0.09, -0.01] & 0.009 \\
Persuasive $\times$ AI-mediated outreach & 0.40 & [0.14, 0.66] & 0.003 \\
RQ1: AI-mediated outreach $\times$ Avoidance of political conversations & 0.02 & [-0.06, 0.10] & 0.627 \\
\addlinespace
Observations & 1800 &  &  \\
$R^2$ / adjusted $R^2$ & 0.032 / 0.030 &  &  \\
\bottomrule
\end{tabular}
\caption{OLS regression model for perceived positive impact of campaign outreach with 95\% confidence intervals, including avoidance of political conversations as a moderator. $N = 1800$.}
\end{table}

\begin{table}[H]
\centering
\begin{tabular}{L{5.8cm}ccc}
\toprule
Predictors & Estimate & 95\% CI & p \\
\midrule
Intercept & 4.49 & [4.42, 4.56] & <0.001 \\
Persuasive outreach & 0.55 & [0.41, 0.69] & <0.001 \\
AI-mediated outreach & 0.49 & [0.35, 0.63] & <0.001 \\
Avoidance of political conversations & 0.16 & [0.12, 0.21] & <0.001 \\
Persuasive $\times$ AI-mediated outreach & -0.37 & [-0.66, -0.09] & 0.009 \\
RQ1: AI-mediated outreach $\times$ Avoidance of political conversations & -0.08 & [-0.16, 0.01] & 0.083 \\
\addlinespace
Observations & 1800 &  &  \\
$R^2$ / adjusted $R^2$ & 0.089 / 0.087 &  &  \\
\bottomrule
\end{tabular}
\caption{OLS regression model for future campaign avoidance with 95\% confidence intervals, including avoidance of political conversations as a moderator. $N = 1800$.}
\end{table}

\begin{table}[H]
\centering
\begin{tabular}{L{5.8cm}ccc}
\toprule
Predictors & Estimate & 95\% CI & p \\
\midrule
Intercept & 4.53 & [4.48, 4.59] & <0.001 \\
Persuasive outreach & 0.40 & [0.29, 0.51] & <0.001 \\
AI-mediated outreach & 0.48 & [0.37, 0.59] & <0.001 \\
Avoidance of political conversations & 0.07 & [0.04, 0.11] & <0.001 \\
Persuasive $\times$ AI-mediated outreach & -0.31 & [-0.53, -0.09] & 0.005 \\
RQ1: AI-mediated outreach $\times$ Avoidance of political conversations & -0.03 & [-0.10, 0.03] & 0.342 \\
\addlinespace
Observations & 1800 &  &  \\
$R^2$ / adjusted $R^2$ & 0.078 / 0.076 &  &  \\
\bottomrule
\end{tabular}
\caption{OLS regression model for penalty for source with 95\% confidence intervals, including avoidance of political conversations as a moderator. $N = 1800$.}
\end{table}

%%%%%%%%%%%%%%%%%%%%%%%%%%%%%%%%%%%%%%%%%%%%%%%%%%%%%%%%%%%%%%%%%%%%

\subsubsection{Feeling toward people with opposing opinions}
\medskip \noindent\textbf{United States}

\begin{table}[H]
\centering
\begin{tabular}{L{5.8cm}ccc}
\toprule
Predictors & Estimate & 95\% CI & p \\
\midrule
Intercept & 3.58 & [3.49, 3.68] & <0.001 \\
Persuasive outreach & -0.23 & [-0.41, -0.04] & 0.017 \\
AI-mediated outreach & 0.03 & [-0.16, 0.21] & 0.781 \\
Feeling toward people with opposing opinions & 0.02 & [0.01, 0.02] & <0.001 \\
Persuasive $\times$ AI-mediated outreach & 0.29 & [-0.08, 0.65] & 0.128 \\
RQ2: AI-mediated outreach $\times$ Feeling toward people with opposing opinions & -0.00 & [-0.01, 0.01] & 0.885 \\
\addlinespace
Observations & 1800 &  &  \\
$R^2$ / adjusted $R^2$ & 0.066 / 0.063 &  &  \\
\bottomrule
\end{tabular}
\caption{OLS regression model for willingness to participate with 95\% confidence intervals, including feeling toward people with opposing opinions as a moderator. $N = 1800$.}
\end{table}

\begin{table}[H]
\centering
\begin{tabular}{L{5.8cm}ccc}
\toprule
Predictors & Estimate & 95\% CI & p \\
\midrule
Intercept & 3.60 & [3.53, 3.68] & <0.001 \\
Persuasive outreach & 0.88 & [0.72, 1.03] & <0.001 \\
AI-mediated outreach & 0.19 & [0.04, 0.34] & 0.016 \\
Feeling toward people with opposing opinions & -0.00 & [-0.01, -0.00] & <0.001 \\
Persuasive $\times$ AI-mediated outreach & -0.17 & [-0.48, 0.13] & 0.264 \\
RQ2: AI-mediated outreach $\times$ Feeling toward people with opposing opinions & -0.00 & [-0.01, 0.00] & 0.199 \\
\addlinespace
Observations & 1800 &  &  \\
$R^2$ / adjusted $R^2$ & 0.075 / 0.073 &  &  \\
\bottomrule
\end{tabular}
\caption{OLS regression model for perceived threat to freedom with 95\% confidence intervals, including feeling toward people with opposing opinions as a moderator. $N = 1800$.}
\end{table}

\begin{table}[H]
\centering
\begin{tabular}{L{5.8cm}ccc}
\toprule
Predictors & Estimate & 95\% CI & p \\
\midrule
Intercept & 3.63 & [3.57, 3.70] & <0.001 \\
Persuasive outreach & -0.52 & [-0.66, -0.39] & <0.001 \\
AI-mediated outreach & -0.58 & [-0.71, -0.45] & <0.001 \\
Feeling toward people with opposing opinions & 0.01 & [0.01, 0.01] & <0.001 \\
Persuasive $\times$ AI-mediated outreach & 0.21 & [-0.05, 0.47] & 0.117 \\
RQ2: AI-mediated outreach $\times$ Feeling toward people with opposing opinions & 0.01 & [0.00, 0.01] & 0.020 \\
\addlinespace
Observations & 1800 &  &  \\
$R^2$ / adjusted $R^2$ & 0.116 / 0.114 &  &  \\
\bottomrule
\end{tabular}
\caption{OLS regression model for acceptability of campaign outreach with 95\% confidence intervals, including feeling toward people with opposing opinions as a moderator. $N = 1800$.}
\end{table}

\begin{table}[H]
\centering
\begin{tabular}{L{5.8cm}ccc}
\toprule
Predictors & Estimate & 95\% CI & p \\
\midrule
Intercept & 3.70 & [3.63, 3.77] & <0.001 \\
Persuasive outreach & -0.34 & [-0.48, -0.20] & <0.001 \\
AI-mediated outreach & -0.44 & [-0.57, -0.30] & <0.001 \\
Feeling toward people with opposing opinions & 0.02 & [0.02, 0.02] & <0.001 \\
Persuasive $\times$ AI-mediated outreach & 0.10 & [-0.17, 0.38] & 0.464 \\
RQ2: AI-mediated outreach $\times$ Feeling toward people with opposing opinions & 0.01 & [0.00, 0.01] & 0.033 \\
\addlinespace
Observations & 1800 &  &  \\
$R^2$ / adjusted $R^2$ & 0.132 / 0.129 &  &  \\
\bottomrule
\end{tabular}
\caption{OLS regression model for perceived positive impact of campaign outreach with 95\% confidence intervals, including feeling toward people with opposing opinions as a moderator. $N = 1800$.}
\end{table}

\begin{table}[H]
\centering
\begin{tabular}{L{5.8cm}ccc}
\toprule
Predictors & Estimate & 95\% CI & p \\
\midrule
Intercept & 4.51 & [4.43, 4.58] & <0.001 \\
Persuasive outreach & 0.44 & [0.29, 0.60] & <0.001 \\
AI-mediated outreach & 0.38 & [0.23, 0.53] & <0.001 \\
Feeling toward people with opposing opinions & -0.01 & [-0.01, -0.01] & <0.001 \\
Persuasive $\times$ AI-mediated outreach & -0.28 & [-0.58, 0.03] & 0.077 \\
RQ2: AI-mediated outreach $\times$ Feeling toward people with opposing opinions & -0.01 & [-0.01, -0.00] & 0.037 \\
\addlinespace
Observations & 1800 &  &  \\
$R^2$ / adjusted $R^2$ & 0.066 / 0.063 &  &  \\
\bottomrule
\end{tabular}
\caption{OLS regression model for future campaign avoidance with 95\% confidence intervals, including feeling toward people with opposing opinions as a moderator. $N = 1800$.}
\end{table}

\begin{table}[H]
\centering
\begin{tabular}{L{5.8cm}ccc}
\toprule
Predictors & Estimate & 95\% CI & p \\
\midrule
Intercept & 4.38 & [4.32, 4.44] & <0.001 \\
Persuasive outreach & 0.40 & [0.28, 0.52] & <0.001 \\
AI-mediated outreach & 0.55 & [0.43, 0.67] & <0.001 \\
Feeling toward people with opposing opinions & -0.01 & [-0.01, -0.01] & <0.001 \\
Persuasive $\times$ AI-mediated outreach & -0.14 & [-0.37, 0.10] & 0.266 \\
RQ2: AI-mediated outreach $\times$ Feeling toward people with opposing opinions & -0.01 & [-0.01, -0.00] & <0.001 \\
\addlinespace
Observations & 1800 &  &  \\
$R^2$ / adjusted $R^2$ & 0.109 / 0.107 &  &  \\
\bottomrule
\end{tabular}
\caption{OLS regression model for penalty for source with 95\% confidence intervals, including feeling toward people with opposing opinions as a moderator. $N = 1800$.}
\end{table}

\medskip \noindent\textbf{United Kingdom}
\begin{table}[H]
\centering
\begin{tabular}{L{5.8cm}ccc}
\toprule
Predictors & Estimate & 95\% CI & p \\
\midrule
Intercept & 3.80 & [3.71, 3.89] & <0.001 \\
Persuasive outreach & -0.24 & [-0.41, -0.06] & 0.007 \\
AI-mediated outreach & -0.01 & [-0.18, 0.17] & 0.931 \\
Feeling toward people with opposing opinions & 0.01 & [0.01, 0.02] & <0.001 \\
Persuasive $\times$ AI-mediated outreach & 0.42 & [0.08, 0.77] & 0.016 \\
RQ2: AI-mediated outreach $\times$ Feeling toward people with opposing opinions & -0.00 & [-0.01, 0.00] & 0.470 \\
\addlinespace
Observations & 1800 &  &  \\
$R^2$ / adjusted $R^2$ & 0.038 / 0.036 &  &  \\
\bottomrule
\end{tabular}
\caption{OLS regression model for willingness to participate with 95\% confidence intervals, including feeling toward people with opposing opinions as a moderator. $N = 1800$.}
\end{table}

\begin{table}[H]
\centering
\begin{tabular}{L{5.8cm}ccc}
\toprule
Predictors & Estimate & 95\% CI & p \\
\midrule
Intercept & 3.81 & [3.74, 3.88] & <0.001 \\
Persuasive outreach & 0.79 & [0.65, 0.93] & <0.001 \\
AI-mediated outreach & 0.26 & [0.12, 0.41] & <0.001 \\
Feeling toward people with opposing opinions & -0.00 & [-0.01, -0.00] & 0.001 \\
Persuasive $\times$ AI-mediated outreach & -0.26 & [-0.54, 0.02] & 0.073 \\
RQ2: AI-mediated outreach $\times$ Feeling toward people with opposing opinions & 0.00 & [-0.00, 0.01] & 0.215 \\
\addlinespace
Observations & 1800 &  &  \\
$R^2$ / adjusted $R^2$ & 0.076 / 0.073 &  &  \\
\bottomrule
\end{tabular}
\caption{OLS regression model for perceived threat to freedom with 95\% confidence intervals, including feeling toward people with opposing opinions as a moderator. $N = 1800$.}
\end{table}

\begin{table}[H]
\centering
\begin{tabular}{L{5.8cm}ccc}
\toprule
Predictors & Estimate & 95\% CI & p \\
\midrule
Intercept & 3.59 & [3.53, 3.65] & <0.001 \\
Persuasive outreach & -0.57 & [-0.69, -0.44] & <0.001 \\
AI-mediated outreach & -0.53 & [-0.65, -0.41] & <0.001 \\
Feeling toward people with opposing opinions & 0.01 & [0.01, 0.01] & <0.001 \\
Persuasive $\times$ AI-mediated outreach & 0.22 & [-0.02, 0.46] & 0.076 \\
RQ2: AI-mediated outreach $\times$ Feeling toward people with opposing opinions & 0.00 & [-0.00, 0.01] & 0.210 \\
\addlinespace
Observations & 1800 &  &  \\
$R^2$ / adjusted $R^2$ & 0.108 / 0.105 &  &  \\
\bottomrule
\end{tabular}
\caption{OLS regression model for acceptability of campaign outreach with 95\% confidence intervals, including feeling toward people with opposing opinions as a moderator. $N = 1800$.}
\end{table}

\begin{table}[H]
\centering
\begin{tabular}{L{5.8cm}ccc}
\toprule
Predictors & Estimate & 95\% CI & p \\
\midrule
Intercept & 3.87 & [3.81, 3.94] & <0.001 \\
Persuasive outreach & -0.31 & [-0.43, -0.18] & <0.001 \\
AI-mediated outreach & -0.28 & [-0.41, -0.16] & <0.001 \\
Feeling toward people with opposing opinions & 0.01 & [0.01, 0.02] & <0.001 \\
Persuasive $\times$ AI-mediated outreach & 0.41 & [0.16, 0.67] & 0.001 \\
RQ2: AI-mediated outreach $\times$ Feeling toward people with opposing opinions & 0.00 & [-0.00, 0.01] & 0.424 \\
\addlinespace
Observations & 1800 &  &  \\
$R^2$ / adjusted $R^2$ & 0.097 / 0.094 &  &  \\
\bottomrule
\end{tabular}
\caption{OLS regression model for perceived positive impact of campaign outreach with 95\% confidence intervals, including feeling toward people with opposing opinions as a moderator. $N = 1800$.}
\end{table}

\begin{table}[H]
\centering
\begin{tabular}{L{5.8cm}ccc}
\toprule
Predictors & Estimate & 95\% CI & p \\
\midrule
Intercept & 4.49 & [4.42, 4.56] & <0.001 \\
Persuasive outreach & 0.56 & [0.42, 0.70] & <0.001 \\
AI-mediated outreach & 0.48 & [0.34, 0.62] & <0.001 \\
Feeling toward people with opposing opinions & -0.01 & [-0.01, -0.01] & <0.001 \\
Persuasive $\times$ AI-mediated outreach & -0.36 & [-0.65, -0.08] & 0.012 \\
RQ2: AI-mediated outreach $\times$ Feeling toward people with opposing opinions & -0.00 & [-0.01, 0.01] & 0.879 \\
\addlinespace
Observations & 1800 &  &  \\
$R^2$ / adjusted $R^2$ & 0.077 / 0.075 &  &  \\
\bottomrule
\end{tabular}
\caption{OLS regression model for future campaign avoidance with 95\% confidence intervals, including feeling toward people with opposing opinions as a moderator. $N = 1800$.}
\end{table}

\begin{table}[H]
\centering
\begin{tabular}{L{5.8cm}ccc}
\toprule
Predictors & Estimate & 95\% CI & p \\
\midrule
Intercept & 4.53 & [4.48, 4.58] & <0.001 \\
Persuasive outreach & 0.40 & [0.29, 0.51] & <0.001 \\
AI-mediated outreach & 0.46 & [0.35, 0.57] & <0.001 \\
Feeling toward people with opposing opinions & -0.01 & [-0.01, -0.01] & <0.001 \\
Persuasive $\times$ AI-mediated outreach & -0.32 & [-0.53, -0.10] & 0.004 \\
RQ2: AI-mediated outreach $\times$ Feeling toward people with opposing opinions & -0.00 & [-0.01, -0.00] & 0.050 \\
\addlinespace
Observations & 1800 &  &  \\
$R^2$ / adjusted $R^2$ & 0.108 / 0.105 &  &  \\
\bottomrule
\end{tabular}
\caption{OLS regression model for penalty for source with 95\% confidence intervals, including feeling toward people with opposing opinions as a moderator. $N = 1800$.}
\end{table}

%%%%%%%%%%%%%%%%%%%%%%%%%%%%%%%%%%%%%%%%%%%%%%%%%%%%%%%%%%%%%%%%%%%%

\subsubsection{AI risk perception}
\medskip \noindent\textbf{United States}

\begin{table}[H]
\centering
\begin{tabular}{L{5.8cm}ccc}
\toprule
Predictors & Estimate & 95\% CI & p \\
\midrule
Intercept & 3.58 & [3.49, 3.67] & <0.001 \\
Persuasive outreach & -0.22 & [-0.41, -0.03] & 0.020 \\
AI-mediated outreach & -0.00 & [-0.19, 0.19] & 0.991 \\
RQ3: AI risk perception & -0.23 & [-0.30, -0.16] & <0.001 \\
Persuasive $\times$ AI-mediated outreach & 0.20 & [-0.17, 0.58] & 0.291 \\
RQ3: AI-mediated outreach $\times$ AI risk perception & -0.15 & [-0.30, -0.01] & 0.034 \\
\addlinespace
Observations & 1800 &  &  \\
$R^2$ / adjusted $R^2$ & 0.027 / 0.024 &  &  \\
\bottomrule
\end{tabular}
\caption{OLS regression model for willingness to participate with 95\% confidence intervals, including AI risk perception as a moderator. $N = 1800$.}
\end{table}

\begin{table}[H]
\centering
\begin{tabular}{L{5.8cm}ccc}
\toprule
Predictors & Estimate & 95\% CI & p \\
\midrule
Intercept & 3.61 & [3.53, 3.68] & <0.001 \\
Persuasive outreach & 0.91 & [0.76, 1.06] & <0.001 \\
AI-mediated outreach & 0.21 & [0.06, 0.36] & 0.006 \\
RQ3: AI risk perception & 0.22 & [0.16, 0.28] & <0.001 \\
Persuasive $\times$ AI-mediated outreach & -0.13 & [-0.43, 0.17] & 0.394 \\
RQ3: AI-mediated outreach $\times$ AI risk perception & 0.14 & [0.02, 0.25] & 0.019 \\
\addlinespace
Observations & 1800 &  &  \\
$R^2$ / adjusted $R^2$ & 0.101 / 0.098 &  &  \\
\bottomrule
\end{tabular}
\caption{OLS regression model for perceived threat to freedom with 95\% confidence intervals, including AI risk perception as a moderator. $N = 1800$.}
\end{table}

\begin{table}[H]
\centering
\begin{tabular}{L{5.8cm}ccc}
\toprule
Predictors & Estimate & 95\% CI & p \\
\midrule
Intercept & 3.63 & [3.56, 3.69] & <0.001 \\
Persuasive outreach & -0.55 & [-0.68, -0.42] & <0.001 \\
AI-mediated outreach & -0.61 & [-0.74, -0.48] & <0.001 \\
RQ3: AI risk perception & -0.29 & [-0.34, -0.24] & <0.001 \\
Persuasive $\times$ AI-mediated outreach & 0.15 & [-0.11, 0.41] & 0.255 \\
RQ3: AI-mediated outreach $\times$ AI risk perception & -0.14 & [-0.24, -0.04] & 0.006 \\
\addlinespace
Observations & 1800 &  &  \\
$R^2$ / adjusted $R^2$ & 0.134 / 0.131 &  &  \\
\bottomrule
\end{tabular}
\caption{OLS regression model for acceptability of campaign outreach with 95\% confidence intervals, including AI risk perception as a moderator. $N = 1800$.}
\end{table}

\begin{table}[H]
\centering
\begin{tabular}{L{5.8cm}ccc}
\toprule
Predictors & Estimate & 95\% CI & p \\
\midrule
Intercept & 3.70 & [3.63, 3.77] & <0.001 \\
Persuasive outreach & -0.34 & [-0.48, -0.20] & <0.001 \\
AI-mediated outreach & -0.47 & [-0.61, -0.32] & <0.001 \\
RQ3: AI risk perception & -0.25 & [-0.30, -0.20] & <0.001 \\
Persuasive $\times$ AI-mediated outreach & 0.04 & [-0.25, 0.32] & 0.791 \\
RQ3: AI-mediated outreach $\times$ AI risk perception & -0.11 & [-0.22, 0.00] & 0.051 \\
\addlinespace
Observations & 1800 &  &  \\
$R^2$ / adjusted $R^2$ & 0.073 / 0.071 &  &  \\
\bottomrule
\end{tabular}
\caption{OLS regression model for perceived positive impact of campaign outreach with 95\% confidence intervals, including AI risk perception as a moderator. $N = 1800$.}
\end{table}

\begin{table}[H]
\centering
\begin{tabular}{L{5.8cm}ccc}
\toprule
Predictors & Estimate & 95\% CI & p \\
\midrule
Intercept & 4.51 & [4.44, 4.59] & <0.001 \\
Persuasive outreach & 0.48 & [0.33, 0.63] & <0.001 \\
AI-mediated outreach & 0.41 & [0.26, 0.56] & <0.001 \\
RQ3: AI risk perception & 0.32 & [0.26, 0.38] & <0.001 \\
Persuasive $\times$ AI-mediated outreach & -0.21 & [-0.51, 0.09] & 0.178 \\
RQ3: AI-mediated outreach $\times$ AI risk perception & 0.20 & [0.08, 0.31] & <0.001 \\
\addlinespace
Observations & 1800 &  &  \\
$R^2$ / adjusted $R^2$ & 0.097 / 0.095 &  &  \\
\bottomrule
\end{tabular}
\caption{OLS regression model for future campaign avoidance with 95\% confidence intervals, including AI risk perception as a moderator. $N = 1800$.}
\end{table}

\begin{table}[H]
\centering
\begin{tabular}{L{5.8cm}ccc}
\toprule
Predictors & Estimate & 95\% CI & p \\
\midrule
Intercept & 4.38 & [4.32, 4.44] & <0.001 \\
Persuasive outreach & 0.43 & [0.31, 0.55] & <0.001 \\
AI-mediated outreach & 0.59 & [0.47, 0.70] & <0.001 \\
RQ3: AI risk perception & 0.30 & [0.26, 0.35] & <0.001 \\
Persuasive $\times$ AI-mediated outreach & -0.08 & [-0.32, 0.15] & 0.483 \\
RQ3: AI-mediated outreach $\times$ AI risk perception & 0.13 & [0.04, 0.22] & 0.004 \\
\addlinespace
Observations & 1800 &  &  \\
$R^2$ / adjusted $R^2$ & 0.152 / 0.149 &  &  \\
\bottomrule
\end{tabular}
\caption{OLS regression model for penalty for source with 95\% confidence intervals, including AI risk perception as a moderator. $N = 1800$.}
\end{table}

\medskip \noindent\textbf{United Kingdom}
\begin{table}[H]
\centering
\begin{tabular}{L{5.8cm}ccc}
\toprule
Predictors & Estimate & 95\% CI & p \\
\midrule
Intercept & 3.80 & [3.72, 3.89] & <0.001 \\
Persuasive outreach & -0.23 & [-0.41, -0.06] & 0.010 \\
AI-mediated outreach & -0.03 & [-0.20, 0.15] & 0.755 \\
RQ3: AI risk perception & -0.16 & [-0.23, -0.08] & <0.001 \\
Persuasive $\times$ AI-mediated outreach & 0.43 & [0.08, 0.78] & 0.017 \\
RQ3: AI-mediated outreach $\times$ AI risk perception & -0.09 & [-0.24, 0.06] & 0.260 \\
\addlinespace
Observations & 1800 &  &  \\
$R^2$ / adjusted $R^2$ & 0.017 / 0.014 &  &  \\
\bottomrule
\end{tabular}
\caption{OLS regression model for willingness to participate with 95\% confidence intervals, including AI risk perception as a moderator. $N = 1800$.}
\end{table}

\begin{table}[H]
\centering
\begin{tabular}{L{5.8cm}ccc}
\toprule
Predictors & Estimate & 95\% CI & p \\
\midrule
Intercept & 3.81 & [3.74, 3.88] & <0.001 \\
Persuasive outreach & 0.78 & [0.64, 0.92] & <0.001 \\
AI-mediated outreach & 0.26 & [0.12, 0.40] & <0.001 \\
RQ3: AI risk perception & 0.23 & [0.17, 0.30] & <0.001 \\
Persuasive $\times$ AI-mediated outreach & -0.29 & [-0.57, -0.01] & 0.045 \\
RQ3: AI-mediated outreach $\times$ AI risk perception & 0.08 & [-0.04, 0.20] & 0.196 \\
\addlinespace
Observations & 1800 &  &  \\
$R^2$ / adjusted $R^2$ & 0.100 / 0.097 &  &  \\
\bottomrule
\end{tabular}
\caption{OLS regression model for perceived threat to freedom with 95\% confidence intervals, including AI risk perception as a moderator. $N = 1800$.}
\end{table}

\begin{table}[H]
\centering
\begin{tabular}{L{5.8cm}ccc}
\toprule
Predictors & Estimate & 95\% CI & p \\
\midrule
Intercept & 3.59 & [3.53, 3.65] & <0.001 \\
Persuasive outreach & -0.56 & [-0.68, -0.44] & <0.001 \\
AI-mediated outreach & -0.54 & [-0.66, -0.42] & <0.001 \\
RQ3: AI risk perception & -0.23 & [-0.28, -0.17] & <0.001 \\
Persuasive $\times$ AI-mediated outreach & 0.24 & [-0.00, 0.48] & 0.051 \\
RQ3: AI-mediated outreach $\times$ AI risk perception & -0.13 & [-0.23, -0.02] & 0.017 \\
\addlinespace
Observations & 1800 &  &  \\
$R^2$ / adjusted $R^2$ & 0.121 / 0.118 &  &  \\
\bottomrule
\end{tabular}
\caption{OLS regression model for acceptability of campaign outreach with 95\% confidence intervals, including AI risk perception as a moderator. $N = 1800$.}
\end{table}

\begin{table}[H]
\centering
\begin{tabular}{L{5.8cm}ccc}
\toprule
Predictors & Estimate & 95\% CI & p \\
\midrule
Intercept & 3.87 & [3.81, 3.94] & <0.001 \\
Persuasive outreach & -0.30 & [-0.43, -0.17] & <0.001 \\
AI-mediated outreach & -0.31 & [-0.43, -0.18] & <0.001 \\
RQ3: AI risk perception & -0.16 & [-0.21, -0.10] & <0.001 \\
Persuasive $\times$ AI-mediated outreach & 0.41 & [0.16, 0.67] & 0.002 \\
RQ3: AI-mediated outreach $\times$ AI risk perception & -0.14 & [-0.25, -0.03] & 0.015 \\
\addlinespace
Observations & 1800 &  &  \\
$R^2$ / adjusted $R^2$ & 0.048 / 0.045 &  &  \\
\bottomrule
\end{tabular}
\caption{OLS regression model for perceived positive impact of campaign outreach with 95\% confidence intervals, including AI risk perception as a moderator. $N = 1800$.}
\end{table}

\begin{table}[H]
\centering
\begin{tabular}{L{5.8cm}ccc}
\toprule
Predictors & Estimate & 95\% CI & p \\
\midrule
Intercept & 4.49 & [4.42, 4.56] & <0.001 \\
Persuasive outreach & 0.55 & [0.41, 0.69] & <0.001 \\
AI-mediated outreach & 0.48 & [0.34, 0.62] & <0.001 \\
RQ3: AI risk perception & 0.26 & [0.20, 0.32] & <0.001 \\
Persuasive $\times$ AI-mediated outreach & -0.39 & [-0.67, -0.11] & 0.007 \\
RQ3: AI-mediated outreach $\times$ AI risk perception & 0.20 & [0.08, 0.32] & 0.001 \\
\addlinespace
Observations & 1800 &  &  \\
$R^2$ / adjusted $R^2$ & 0.098 / 0.096 &  &  \\
\bottomrule
\end{tabular}
\caption{OLS regression model for future campaign avoidance with 95\% confidence intervals, including AI risk perception as a moderator. $N = 1800$.}
\end{table}

\begin{table}[H]
\centering
\begin{tabular}{L{5.8cm}ccc}
\toprule
Predictors & Estimate & 95\% CI & p \\
\midrule
Intercept & 4.53 & [4.48, 4.58] & <0.001 \\
Persuasive outreach & 0.39 & [0.29, 0.50] & <0.001 \\
AI-mediated outreach & 0.47 & [0.36, 0.58] & <0.001 \\
RQ3: AI risk perception & 0.24 & [0.19, 0.28] & <0.001 \\
Persuasive $\times$ AI-mediated outreach & -0.34 & [-0.55, -0.12] & 0.002 \\
RQ3: AI-mediated outreach $\times$ AI risk perception & 0.17 & [0.08, 0.27] & <0.001 \\
\addlinespace
Observations & 1800 &  &  \\
$R^2$ / adjusted $R^2$ & 0.123 / 0.120 &  &  \\
\bottomrule
\end{tabular}
\caption{OLS regression model for penalty for source with 95\% confidence intervals, including AI risk perception as a moderator. $N = 1800$.}
\end{table}

\subsection{Manipulation check and instrumental variable robustness check}
\label{section:manipchecks}
Most respondents remembered that the campaign’s stance opposed their own position (US: 90.7\%; UK: 91.4\%). However, recognition of the specific treatment condition was lower in the human-mediated and informational conditions than in the AI-mediated and persuasive conditions. This pattern is plausible for two reasons. First, the detailed manipulation-check items were asked only at the very end of the survey, after respondents had already completed a large number of outcome measures, which likely made exact recall more difficult. Second, even outreach designed to be informational may still be perceived as somewhat persuasive, making this distinction less clear-cut for respondents.

As a robustness check, we therefore used an instrumental variable (IV) approach \citeSI{montgomery_how_2018} to estimate treatment effects for participants whose manipulation-check responses aligned with their assigned condition. Random treatment assignment served as the instrument for perceived treatment.

Compliance with the source manipulation was high in the AI condition (US: 95.2\%; UK: 97.6\%) and moderate in the human condition (US: 65.1\%; UK: 61.0\%). Compliance with the outreach manipulation was high in the persuasive condition (US: 85.2\%; UK: 87.2\%) and moderate in the informational condition (US: 60.0\%; UK: 54.3\%). Importantly, participants who answered ``not sure / can’t remember'' were coded as non-compliers, alongside those who selected the incorrect answers. This makes the IV analysis conservative, as some of these responses may reflect uncertainty or imperfect recall rather than complete failure to register the treatment.

The IV estimates were consistent with the intent-to-treat results in terms of statistical significance and the overall pattern of effects across all six outcomes in both countries. These findings indicate that the main results are not driven by participants who failed the manipulation checks. Because this specification uses the human-mediated and informational conditions as the focal categories, the IV coefficients have the opposite sign relative to the main specification, while implying the same substantive conclusions.

\begin{table}[H]
\centering
\label{tab:iv_reversed_specs}
\footnotesize
\setlength{\tabcolsep}{4pt}
\renewcommand{\arraystretch}{1.1}
\begin{tabular}{L{3.6cm}L{5.2cm}L{5.2cm}}
\toprule
Outcome & Human-focused: Human-mediated outreach ($d_{\text{human}}$) & Informational-focused: Informational outreach ($d_{\text{info}}$) \\
\midrule
H1a/H2a: Willingness to participate 
& $-0.04$ [$-0.34$, $0.26$], $p=.813$ 
& $0.35$ [$-0.01$, $0.72$], $p=.060$ \\

H1b/H2b: Perceived threat to freedom 
& $-0.30$ [$-0.54$, $-0.06$], $p=.016$ 
& $-1.69$ [$-1.99$, $-1.39$], $p<.001$ \\

H1c/H2c: Acceptability 
& $0.92$ [$0.70$, $1.14$], $p<.001$ 
& $0.97$ [$0.71$, $1.23$], $p<.001$ \\

H1d/H2d: Perceived positive impact 
& $0.69$ [$0.46$, $0.92$], $p<.001$ 
& $0.58$ [$0.30$, $0.86$], $p<.001$ \\

H1e/H2e: Future campaign avoidance 
& $-0.60$ [$-0.85$, $-0.35$], $p<.001$ 
& $-0.82$ [$-1.12$, $-0.52$], $p<.001$ \\

H1f/H2f: Penalty for source 
& $-0.87$ [$-1.07$, $-0.68$], $p<.001$ 
& $-0.74$ [$-0.97$, $-0.50$], $p<.001$ \\
\bottomrule
\end{tabular}

\vspace{0.5ex}
\begin{minipage}{0.97\linewidth}
\footnotesize
\caption{Alternative IV specifications using reversed focal categories for the US sample. Entries report unstandardized coefficients with 95\% confidence intervals and two-sided $p$-values from alternative IV specifications using reversed focal categories. }
\end{minipage}
\end{table}

\section{Content analysis of issues}
\label{section:issues}
In our study, we asked participants to name an issue that is important to them using the following question: "People have different political priorities. Please think of one political or social issue that matters to you personally. Briefly name the issue in the field below."

\begin{figure}[H]
    \centering
    \includegraphics[width=\linewidth]{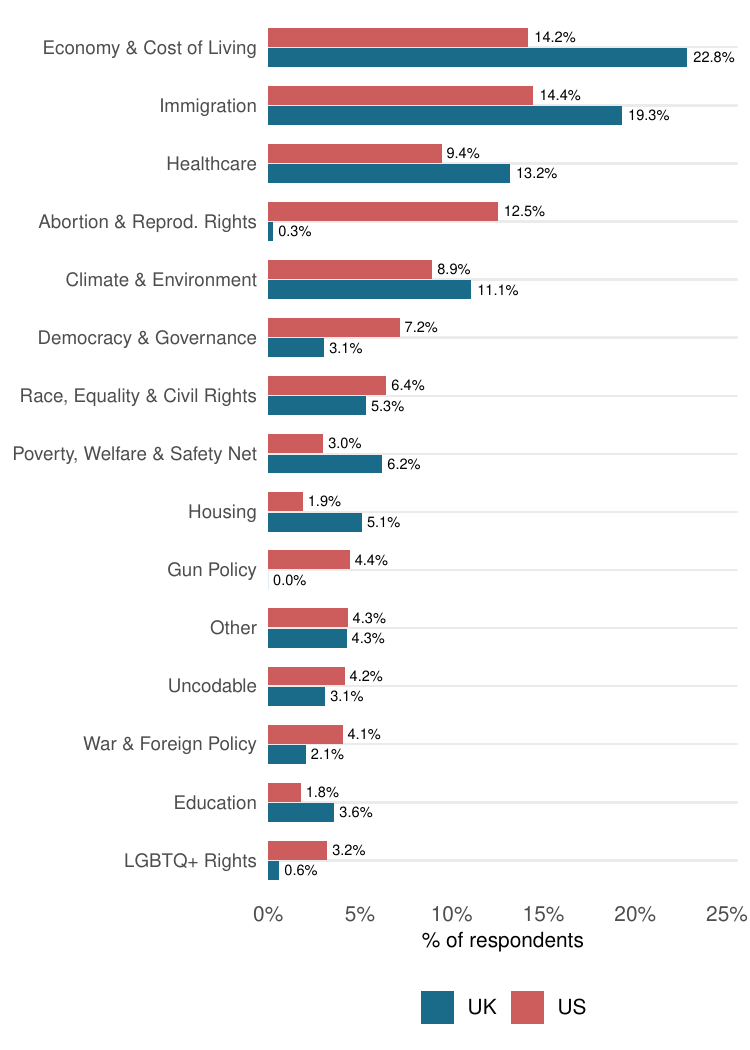}
    \caption{Distribution of issue categories in the US and the UK.}
    \label{fig:issues}
\end{figure}

\subsection{Coding procedure}
In an additional analysis, we classified these open-ended responses into broader issue categories. We first used Claude Code (Sonnet 4.6) to inspect all responses and propose 10--15 broader categories into which the issues could be grouped. Claude Code generated 15 candidate categories, along with category definitions and a classification prompt. We then manually reviewed these categories and made minor revisions to the prompt (see Section~\ref{sec:issue_prompt}). To validate the resulting coding scheme, the two authors of the study independently manually coded a random sample of 50 responses. The same responses were also classified automatically using OpenAI's \texttt{gpt-4o-mini-2024-07-18} model with the temperature set to 0. The two human coders and the automatic classifier showed good intercoder reliability (Krippendorff's alpha = 0.81). After this validation step, we used the model to classify all remaining responses.

\subsection{Issue distributions}
\begin{figure}[H]
    \centering
    \includegraphics[width=\linewidth]{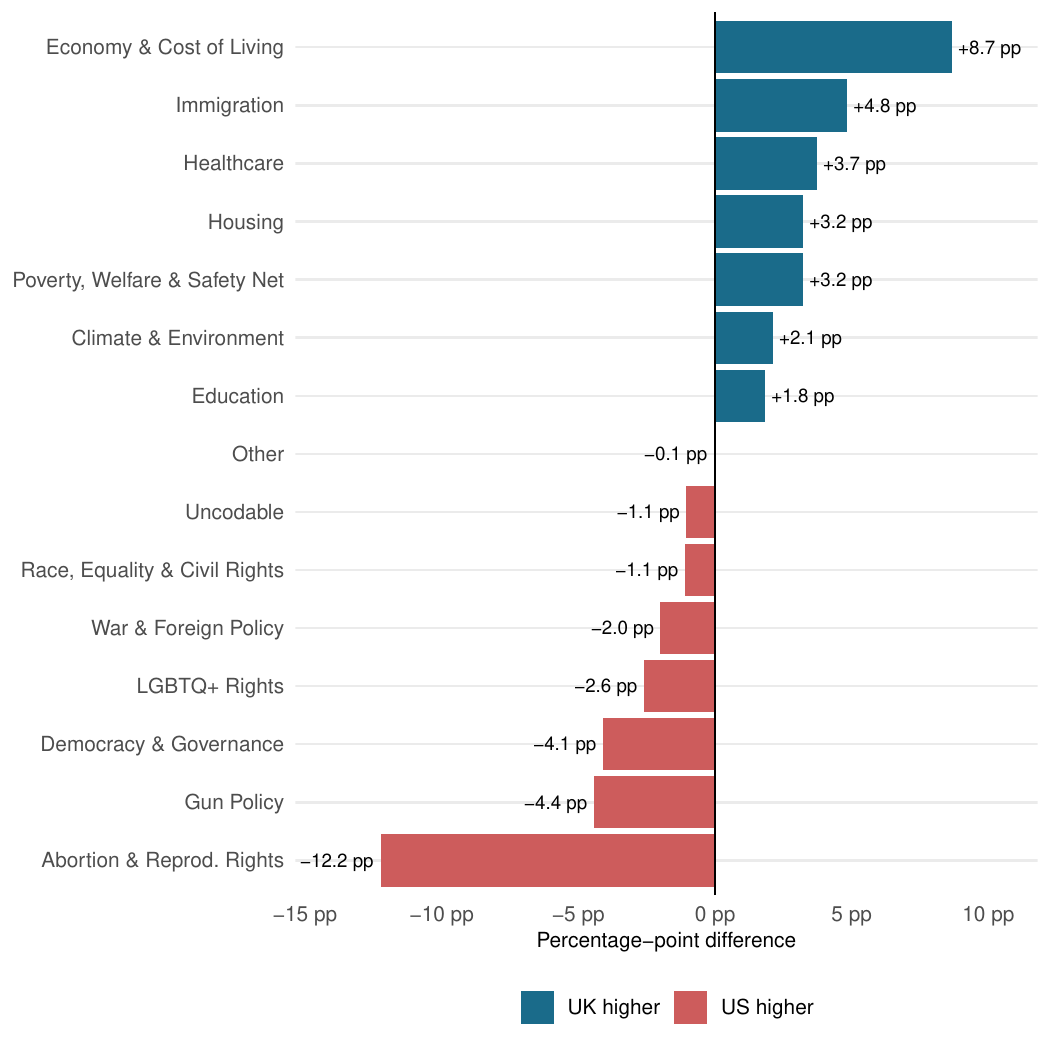}
    \caption{Percentage-point difference (UK--US); positive values indicate that an issue was more common in the UK.}
    \label{fig:issues_over}
\end{figure}

\begin{figure}[H]
    \centering
    \includegraphics[width=\linewidth]{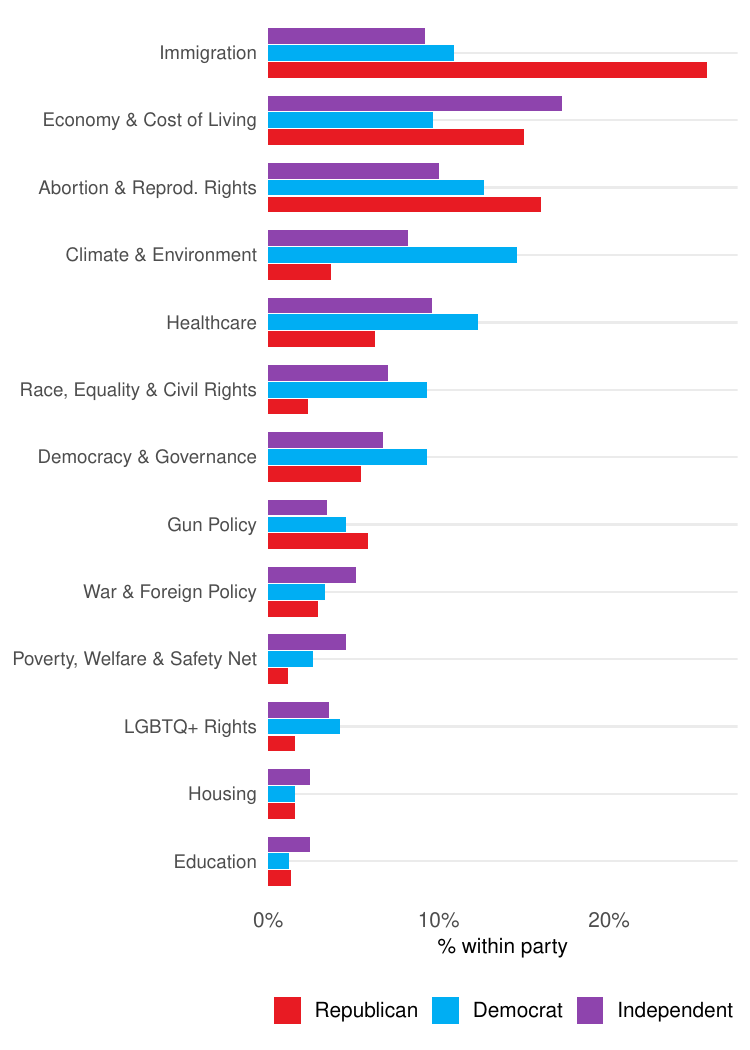}
    \caption{Party identification composition of issue salience for the US sample. Bars show the percentage of each party's respondents who mention a given issue category. Uncodable and other categories are not shown.}
    \label{fig:issues_us}
\end{figure}

\begin{figure}[H]
    \centering
    \includegraphics[width=\linewidth]{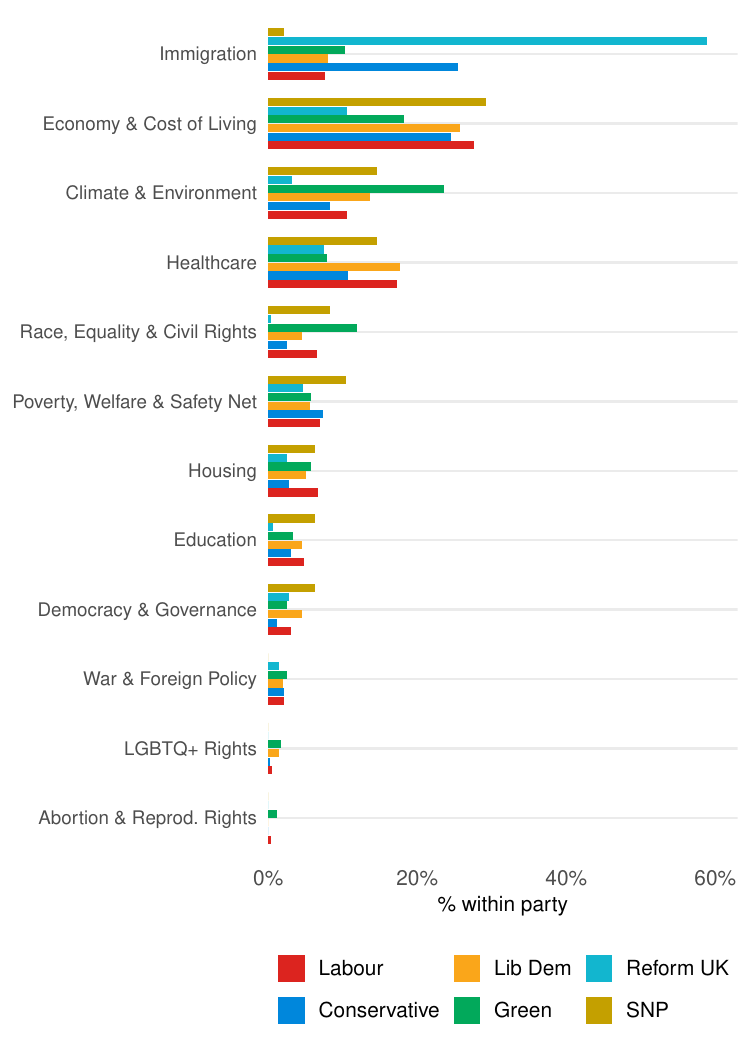}
    \caption{Party identification composition of issue salience for the UK sample. Bars show the percentage of each party's respondents who mention a given issue category. Uncodable and other categories are not shown.}
    \label{fig:issues_uk}
\end{figure}

Overall, the economy and cost of living, immigration, healthcare, and climate change were the most frequently mentioned issues in both countries (see Figure~\ref{fig:issues}). Furthermore, the issue distributions in each country clearly reflect the respective political context. For example, abortion and gun policy were clearly overrepresented in the US and largely absent in the UK, in contrast to the cost of living and healthcare, which were mentioned relatively more often in the UK (see Figure~\ref{fig:issues_over}). When analyzed alongside participants' partisan identification, the issue profiles also broadly aligned with party cleavages in both the US (see Figure~\ref{fig:issues_us}) and the UK (see Figure~\ref{fig:issues_uk}).

\subsection{Classification prompt}
\label{sec:issue_prompt}
We used the following prompt with OpenAI's model:
\begin{promptbox}
You are a political science research assistant coding open-ended survey responses.
Each response is a participant's answer to: 'What is one political or social issue that matters to you personally?'

Assign ONE primary category from this list:
ECO - Economy & Cost of Living (inflation, wages, taxes, wealth inequality, cost of living)
IMM - Immigration (legal/illegal immigration, border security, ICE, deportation, asylum)
HLT - Healthcare (NHS, universal healthcare, costs, access, mental health, social care)
ABR - Abortion & Reproductive Rights (abortion, pro-life/choice, reproductive rights, bodily autonomy)
ENV - Climate Change & Environment (climate, global warming, pollution, renewables, nature)
GUN - Gun Policy (gun control, gun rights, Second Amendment, firearms)
HSG - Housing (housing crisis, affordability, homelessness as housing failure)
WLF - Poverty, Welfare & Social Safety Net (poverty, benefits, pensions, food insecurity, student debt, disability support)
LGB - LGBTQ+ Rights (gay/trans rights, same-sex marriage, gender identity)
EQL - Race, Equality & Civil Rights (racism, civil rights, women's rights, discrimination, social justice)
EDU - Education (school/university quality, funding, tuition, curriculum)
DEM - Democracy, Governance & Rule of Law (democracy, voting, corruption, free speech, political reform, far right)
WAR - War, Security & Foreign Policy (war, Ukraine, Iran, Gaza, defence, national security)
OTH - Other political/social issue not listed above (e.g., crime/public safety, AI regulation, animal welfare, drug policy)
UNK - Uncodable (vague, blank, non-political, or partisan without issue)

Rules:
- Assign primary = the most prominent issue in the response
- If the response names no recognisable issue, assign UNK
- Only return the three letter label and nothing else
Here is the issue that you should classify: 
\end{promptbox}

\section{Additional analysis with political orientation}
\label{section:polor}
\begin{figure}[H]
    \centering
    \includegraphics[width=\linewidth]{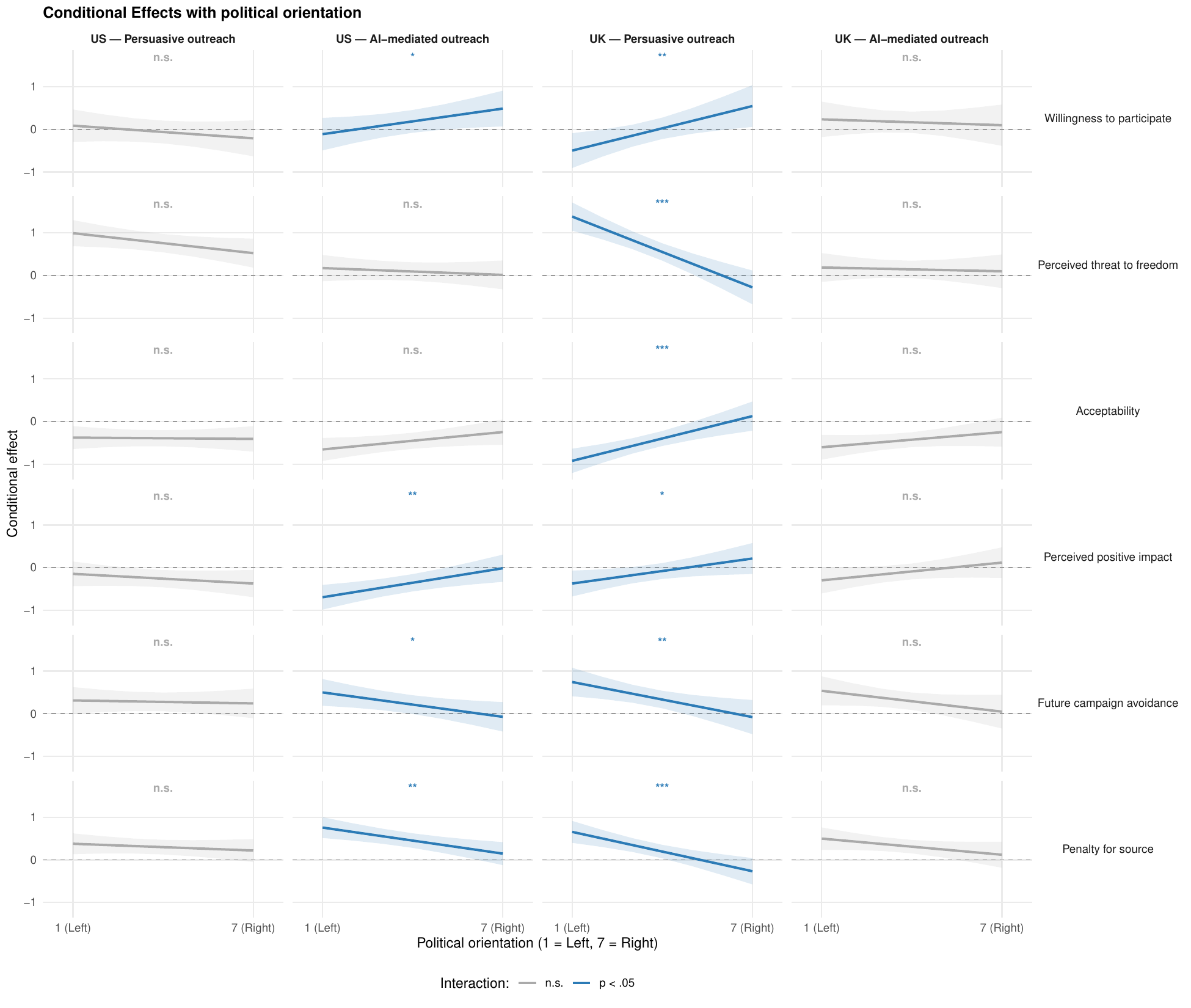}
\caption{Conditional effects by political orientation in the UK and the US.}
    \label{fig:polor}
\end{figure}

In both countries, we measured political orientation on a seven-point scale (1 = left, 7 = right). As the treatment effects may vary by political orientation, we also examined interactions between political orientation and our main independent variables. Political orientation is a significant moderator of human- vs. AI-mediated outreach for some dependent variables. In the UK, by contrast, it is a significant moderator of informational- vs. persuasive outreach for all dependent variables (see Figure~\ref{fig:polor}). In the UK, the more left-leaning respondents are, the stronger the effect of persuasive outreach. In the US, by contrast, the more left-leaning respondents are, the stronger the effect of AI-mediated outreach. At the same time, these interaction patterns mainly affect the magnitude of the effects rather than their overall direction. Even where political orientation significantly moderates the treatment effects, the substantive pattern remains similar across the ideological spectrum.

\section{Additional analysis with norm violation item}
\label{section:norm}
 As an additional non-preregistered analysis of the normative violation mechanism, we examined the item "This kind of outreach is not how campaigns should operate", one of three items comprising the acceptability composite, separately. The wording of this item explicitly captures normative appropriateness judgments about communicative practice rather than affective evaluation or perceived effectiveness. Consistent with the normative violation account, the AI penalty on this item substantially exceeded the persuasion penalty in both countries (US: AI Cohen's $d$ = 0.37 vs. persuasion Cohen's $d$ = 0.21; UK: AI Cohen's $d$ = 0.36 vs. persuasion Cohen's $d$ = 0.28). This contrasts directly with the pattern for perceived threat to freedom, where the persuasion penalty substantially dominated the AI penalty (US: persuasion Cohen's $d$ = 0.53 vs. AI Cohen's $d$ = 0.11; UK: persuasion Cohen's $d$ = 0.51 vs. AI Cohen's $d$ = 0.17). The reversal across these two indicators, each mechanism producing its largest signal on a different outcome, is consistent with distinct primary pathways: reactance to persuasive intent driving freedom-threat perceptions, and normative concerns about AI as a communicative agent driving appropriateness judgments. These findings do not mean that the two mechanisms are fully separate. Instead, they partly overlap: the AI penalty also increases perceived threat to freedom, and the persuasion penalty also increases agreement with the norm violation item. The double dissociation, therefore, speaks to the relative strength of each pathway, not to an exclusive link between the mechanism and the indicator.

\bibliographystyleSI{apacite}
\bibliographySI{disinformation}

\end{document}